\documentclass[preprint,12pt]{elsarticle}

\usepackage{amssymb}

\journal{Nuclear Inst. and Methods in Physics Research, A }

\usepackage{multirow} 
\usepackage{rotating}
\usepackage{mathrsfs}
\usepackage{booktabs}
\usepackage{longtable}
\usepackage{placeins}
\usepackage{amssymb}
\usepackage{upgreek}
\usepackage{svg}
\usepackage{comment}
\usepackage{pdfpages}
\usepackage{siunitx}
\usepackage{lineno}
\usepackage{float}
\usepackage{lineno}
\usepackage{tikz}
\usepackage[percent]{overpic}
\usepackage{subcaption}
\title{\boldmath Measurements of a LYSO crystal array from threshold to 100\,MeV}
 
\author[1]{O.~Beesley}
\author[2]{J.~Carlton}
\author[3]{B.~Davis-Purcell}
\author[4]{D.~Ding}
\author[2]{S.~Foster}
\author[5]{K.~Frahm}
\author[6]{L.~Gibbons}
\author[2]{T.~Gorringe}
\author[1]{D.W.~Hertzog}
\author[5]{S.~Hochrein}
\author[7]{J.~Hui}
\author[1]{P.~Kammel}
\author[1]{J.~LaBounty}
\author[7,8]{J.~Liu}
\author[1]{R.~Roehnelt}
\author[1]{P.~Schwendimann}
\author[5]{A.~Soter}
\author[1]{E.~Swanson}
\author[1]{B.~Taylor}

\affiliation[1]{Center for Nuclear Physics and Astrophysics (CENPA), University of Washington, Seattle, WA 98195, USA}
\affiliation[2]{University of Kentucky, USA}
\affiliation[3]{TRIUMF, Canada}
\affiliation[4] {Shanghai Institute of Ceramics, Chinese Academy of Sciences (SICCAS) Shanghai, 201899, China}
\affiliation[5]{ETH Zurich, Switzerland}
\affiliation[6]{Cornell University, USA}
\affiliation[7]{Shanghai Jiao Tong University,  Shanghai, 200240, China}
\affiliation[8]{New Cornerstone Science Laboratory, Tsung-Dao Lee Institute, Shanghai Jiao Tong University, Shanghai, 201210, China}

\newcommand{\pie}{$\pi\to e \nu$}
\newcommand{\pme}{$\pi\to\mu\to e$}

\begin{document}

\begin{frontmatter}

\begin{abstract}
We report measurements of ten 
custom-made high-homogeneity LYSO crystals.  The investigation is motivated by the need for a compact, high-resolution, and fast electromagnetic calorimeter for a new rare pion decay experiment.  Each $2.5\times 2.5 \times 18$\,cm$^3$ crystal was first characterized for general light yield properties and then its longitudinal response uniformity and energy resolution were measured using low-energy gamma sources. The ten crystals were assembled as an array and subjected to a 30 - 100\,MeV positron beam with excellent momentum definition. The energy and timing resolutions were measured as a function of energy, and the spatial resolution was determined at 70\,MeV. An additional measurement using monoenergetic 17.6\,MeV gammas produced through a p-Li resonance was later made after the photosensors used in positron testing were improved. As an example of the results, the energy resolution at 70\,MeV of (1.52 $\pm$ 0.03)\,\% is more than two times better than reported results using previous generation LYSO crystals.

\end{abstract}


\end{frontmatter}

\section{Introduction}
\label{sec:intro}

\begin{figure}[h]
    \centering
    \includegraphics[width=0.8\textwidth]{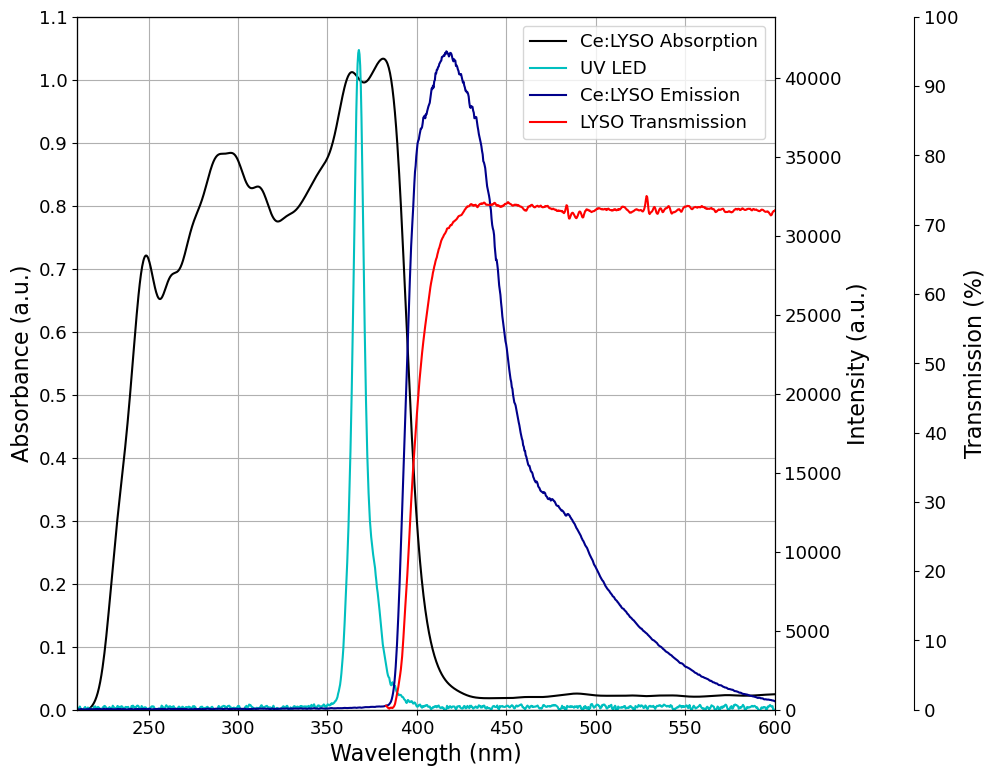}
    \caption{The wavelength dependent optical properties of LYSO. Absorption (black curve, left axis) and transmission (red curve, outer right axis) spectra were measured using an Ocean Optics PX-2 Xenon Light Source to illuminate a LYSO crystal (18 cm length) at one end and a photosensor to read out photons transmitted through the crystal at the other end. The emission spectrum (blue curve, inner right axis) was measured using a 365 nm UV LED (peak absorption wavelength of the cerium dopant in LYSO, cyan curve) to induce scintillation in the crystal.}
    \label{fig:spectroscopy}
\end{figure}

We report on our investigations of the properties of scintillating LYSO crystals as a candidate for the required compact, high-resolution electromagnetic calorimeter for the new PIONEER rare pion decay experiment~\cite{PIONEER:2022yag}. 
LYSO -- lutetium–yttrium oxyorthosilicate -- is typically composed by weight of Lutetium (73\%), Oxygen (18\%), Silicon (6\%), Yttrium (3\%), and a Cerium scintillation dopant ($<0.06\%$). At a density of 7.4\,g/cm$^3$, the radiation length is $X_0 = 1.14$\,cm and the Moli\`{e}re radius is $R_M = 2.07$\,cm.  The light yield of approximately 32,000 photons per MeV\footnote{As measured by SICCAS at the production level}, with a peak wavelength of 420\,nm, rivals NaI(Tl), but at a much faster 40\,ns scintillating decay time. The absorption, emission, and transmission of LYSO crystals is shown in Figure \ref{fig:spectroscopy}. LYSO is both radiation hard and not hygroscopic, which are practical advantages.  Small LYSO crystals are  commonly used in positron-electron-tomography (PET) array instruments (see, e.g., \cite{Bass:2023dmv}). The back-to-back 511\,keV gammas from an $e^{+}e^{-}$ annihilation can each be well resolved and measured with very good time resolution. Such features, exhibited at these low energies, suggest that larger and deeper crystal arrays will provide very good energy and timing resolution at higher energies~\cite{Chen:2007bb,Zhang:2014eww,Mao:2012dr}. 

Despite these attractive properties and successes in PET, use of LYSO crystals in particle physics has been limited. Two relevant experiences involve experiments aiming to measure the monoenergetic 105\,MeV electron from the charged lepton flavor violating (cLFV) muon-to-electron conversion in the field of a nucleus. 
The COMET experiment at J-PARC~\cite{COMET:2018auw} and the Mu2e experiment at Fermilab~\cite{Mu2e:2014fns} both include a calorimeter used to trigger candidate high-energy electrons following muons that stop in a target material. The electrons are registered by the calorimeter and their position is determined by a series of precision tracking wire chambers. The required energy resolution target at 105\,MeV is $5\%$ or better. 
COMET will employ 1920 $2 \times 2 \times 12$\,cm$^3$ (10.5\,$X_0$) LYSO crystals to build a trigger wall. In test beam measurements, they quote a $4.4\%$ resolution at 105\,MeV~\cite{COMET:2018auw} for a small prototype, which meets their requirements. Mu2e originally selected LYSO and tested a $5 \times 5$ array of $3 \times 3 \times 13$\,cm$^3$ (11.2\,$X_0$) crystals. The 4.0\% resolution they achieved also met their specifications~\cite{Atanov:2016blu,7454849}, but the rising cost of the Lutetium Oxide (Lu$_2$O$_3$) -- a required component of LYSO production -- made the calorimeter prohibitively expensive. Consequently, they opted for pure CsI with somewhat reduced resolution~\cite{Atanov:2018ich}.  

Given these experiences, and the intrinsic light yield of LYSO, the question arises as to what might be the ultimate performance of optimized LYSO crystals\footnote{Here, we imply tuning the Ce doping process in the production to achieve high light yield and crystal longitudinal uniformity.}. We ask this in the context of the need for a high-performance calorimeter for positron energies up to 70\,MeV.

\subsection{Calorimetry needed for PIONEER}
The first phase of the newly approved rare pion decay experiment, PIONEER~\cite{PIONEER:2022yag},  aims to measure the branching ratio
\begin{equation}
R_{e/\mu} = \Gamma(\pi^+\rightarrow e^+\nu(\gamma))/\Gamma(\pi^+\rightarrow \mu^+\nu(\gamma))
\end{equation}
to a precision of $\sim 10^{-4}$. This would represent the world's best test of lepton flavor universality. Our experimental design requires a high-resolution and fast electromagnetic calorimeter covering a large solid angle around a highly segmented, active pion stopping target. Positrons from the $\sim 1.25 \times 10^{-4}$ suppressed \pie\ decay are monoenergetic at $E_{e} = m_\pi/2 = 69.3$\,MeV. This signal must be distinguished from the 99.99\% dominant \pme\ decay sequence, where the final 3-body muon decay will result in the Michel positron spectrum with its upper edge at $E_{e} = m_\mu/2 = 53$\,MeV. Building a calorimeter that operates well in this low-energy regime is challenging. The considered options include a monolithic (unsegmented) sphere of liquid xenon (LXe). The MEG Collaboration~\cite{MEGII:2023fog} has demonstrated excellent in-its-class energy and position resolution for a 53\,MeV gamma that would emerge from the cLFV decay $\mu \to e\gamma$ using a LXe calorimeter. An alternative design is based on tapered LYSO crystal segments that together form a truncated polyhedron geometry. 
In principle, LYSO should provide an energy resolution similar to LXe (2\%), since its light yield and signal speed are very similar to LXe, but it has not yet been demonstrated.


In this report, we describe our results from testing ten $2.5 \times 2.5 \times 18$\,cm$^3$ (15.7\,$X_0$) LYSO crystals that were produced at the Shanghai Institute of Ceramics, Chinese Academy of Science (SICCAS). Each of the crystals was evaluated with radioactive sources and then placed, as an array, in test beams to measure their responses from 17 - 100\,MeV. The $\pi$M1 beam line of the High Intensity Proton Accelerator Facility (HIPA) at the Paul Scherrer Institute (PSI) was used to test the array with 30 - 100\,MeV positrons. A 17.6 MeV gamma beam at the Center for Nuclear Physics and Astrophysics at the University of Washington (CENPA) was used to further characterize the array.

\section{Single Crystal Bench Tests}
\label{sec:benchtests}

To characterize the energy resolution of an array, it is first necessary to understand the resolution and longitudinal uniformity of its constituent crystals. Each of the ten LYSO crystals was therefore characterized using radioactive sources at both Shanghai Jiao Tong University (SJTU) and CENPA. The crystals were fabricated at SICCAS with polished surfaces 
and were grown from four different seeds. Each seed initiated a conical ingot from which up to four crystals could be machined. Individual crystal performance was found to be independent of its original seed. Previous generations of LYSO crystals have suffered from non-uniform cerium doping during their growth using the Czochralski method \cite{LYSO:growth}. 
In these crystals, the cerium concentration is lowest at the top of the crystal where the ingot was pulled from the melt first and the concentration rises to a maximum at the point pulled from the melt last.
This cerium concentration non-uniformity resulted in crystal performance that was dependent on the orientation of the crystal with respect to its growth direction. This orientation dependence was not observed for any of the SICCAS LYSO crystals tested for PIONEER.

\subsection{Bench test setups}
An initial evaluation of all ten crystals was performed at SJTU immediately after the production was complete.
The 2.615\,MeV gamma from a Th-232 source was used as a benchmark to study light yield and energy resolution versus impact position along the longitudinal length. Measurements were performed with a source directed at each of the four rectangular ($2.5 \times 18$ cm$^2$) faces; no differences between measurements were observed. The results showed consistent uniformity and energy resolution. These measurements served as an overall acceptance test, after which the 10 crystals were sent to CENPA for additional measurements.


At CENPA, energy resolution measurements were made using three radioactive sources: Na-22, Co-60, and Na-24.\footnote{Na-24 was produced at the CENPA Van de Graaff using the aluminum neutron capture reaction $\text{Al-27} + n \rightarrow \text{Na-24} + \alpha.$} Each of the three sources produces time coincident gamma rays of two separate energies; a third energy can thus be achieved when gamma rays are emitted in a nearly co-linear direction and both are absorbed by the LYSO crystal. The gamma ray energies and their coincident energy is tabulated in Table \ref{tab:gammaEnergies}. An energy resolution measurement was made with each radioactive source at the center position of the longitudinal face of a LYSO crystal, which was wrapped in a 3M Enhanced Specular Reflector Film (ESR). The crystal was coupled using Eljen Technology EJ-550 ($n$ = 1.46)  optical grease to a Hamamatsu R329-02 photo-multiplier tube with a 46 mm diameter photocathode (complete coverage of LYSO square face) operated at -1750\,V. The response was recorded by a $\mu$TCA-based waveform digitizer at a sampling rate of 800 MSPS with a 2\,V dynamic range and 12-bit resolution. In this setup, approximately 1300 photoelectrons were recorded per MeV.
\vskip 1 em
\begin{table}[]
    \centering
    \begin{tabular}{|c|c|c|}
      \hline
      Source Isotope & Gamma Energies [MeV] & Coincident Energy [MeV] \\
      \hline
      Na-22 & 0.511, 1.274 & 1.785 \\
      Co-60 & 1.172, 1.332 & 2.504 \\
      Na-24 & 1.369, 2.754 & 4.123 \\
      \hline
    \end{tabular}
    \caption{Gamma sources and corresponding energies used in single LYSO crystal tests of energy resolution.}
    \label{tab:gammaEnergies}
\end{table}
\vskip 1 em

\subsection{Single crystal energy resolution}
\begin{figure}[H]
    \centering
    \includegraphics[width=0.7\textwidth]{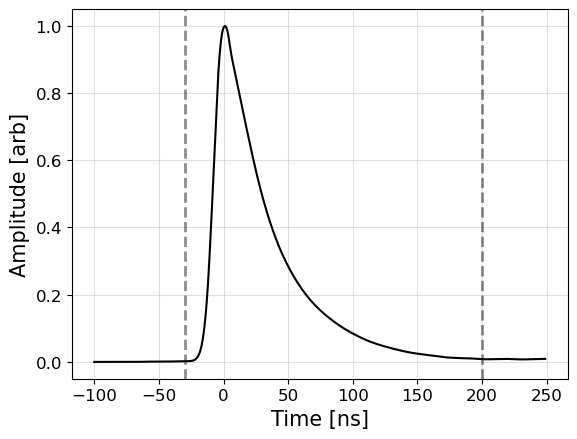}
    \caption{The average waveform response of a LYSO crystal to a 1.17 MeV gamma ray from a Co-60 source. A rise time of 5 ns and a decay time of 40 ns were measured for the crystal. Dashed vertical lines are used to indicate waveform integration bounds which extend from 30 ns before the waveform peak to 200 ns after the peak.}
    \label{fig:avgWF}
\end{figure}

\begin{figure}[H]
    \centering
    \includegraphics[width=0.7\textwidth]{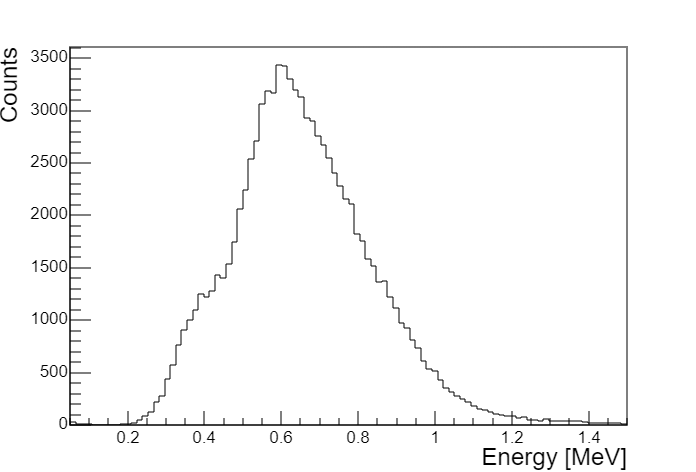}
    \caption{The intrinsic radioactivity energy spectrum of a $2.5 \times 2.5 \times 18$\,cm$^3$ SICCAS LYSO crystal. This radioactivity is produced from Lu-176 beta decay and three subsequent $\gamma$ emissions (307, 202, 88 keV) and is peaked at 0.6 MeV which is the summed energy of the three $\gamma$-rays. }
    \label{fig:intrinsic_LYSO}
\end{figure}

\begin{figure}[H]
    \centering
    \includegraphics[width=0.75\textwidth]{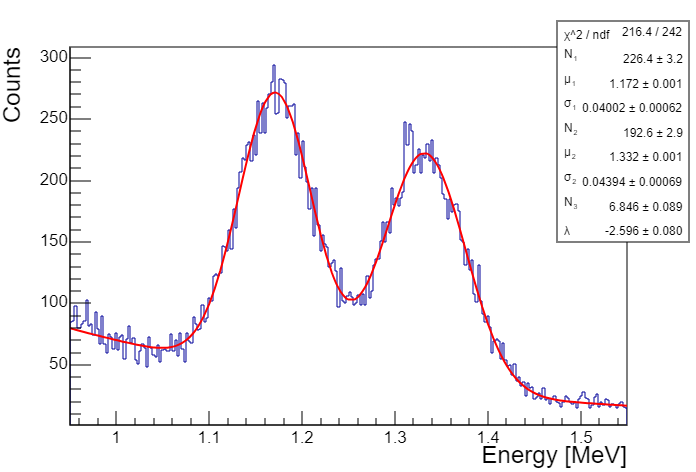}
    \caption{Energy spectrum of 1.17 and 1.33 MeV $\gamma$-rays in a LYSO crystal wrapped in ESR. The two peaks are fit using Gaussian distributions ($\mu_{i},\sigma_{i}$) on an exponential ($\lambda$) background with normalizations $N_{i}$. Peak positions and peak widths are extracted from the fit and are used to compute the energy resolutions at each $\gamma$-ray energy.}
    \label{fig:Co-60-Resolution}
\end{figure}

The waveform response from an energy deposit was integrated using a window $[t_{\mu} - 30, t_{\mu} + 200]$ (ns) where $t_{\mu}$ is the maximal value of the waveform. An averaged waveform response to a Co-60 source with integration bounds superimposed is shown in Figure~\ref{fig:avgWF}. Each LYSO crystal had approximately $37$ \si{kBq} of intrinsic radioactivity resulting mainly from the Lu-176 beta decay and subsequent triple gamma emission. The LYSO radioactivity spectrum is peaked at 0.6 \si{MeV}, but is diffuse and extends from 0 - 1.2 \si{MeV} as shown in Figure~\ref{fig:intrinsic_LYSO}. When radioactive sources of approximately 200 \si{kBq} were placed directly on the crystal surface, pulses above 1 MeV from Lu-176 beta decay were rarely recorded and contributed minimally to the energy resolution. The integrated response of 25,000 Co-60 $\gamma$-ray events is shown in Figure~\ref{fig:Co-60-Resolution}. The resulting distribution can be fit as a sum of an exponential background component and two Gaussian distributions whose peak energies $E_{\mu i}$ and standard deviations $\sigma_{E i}$ are used to calculate energy resolutions $\sigma_{E i} / E_{\mu i}$.

\begin{figure}[H]
    \centering
    \includegraphics[width=0.6\textwidth]{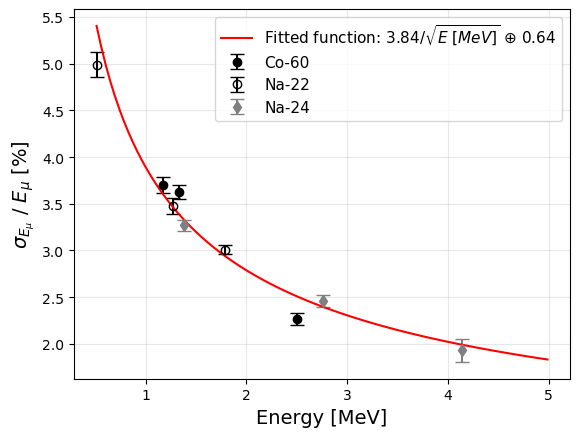}
    \caption{Energy resolution measurements of an ESR-wrapped LYSO crystal at nine different energies obtained using three radioactive $\gamma$-ray sources. A fit of energy resolution as a function of energy is done using the sum of a stochastic term, a noise term, and a constant term. This fit finds no contribution from the noise term and a small contribution from the constant term.} 
    \label{fig:SingleXtalResolutions}
\end{figure}

Energy resolution measurements were performed using all three radioactive sources, which provided nine different energies in total. The resulting energy resolution of a LYSO crystal as a function of $\gamma$-ray energy is shown in Figure~\ref{fig:SingleXtalResolutions}. An electromagnetic calorimeter energy resolution is typically represented with the following functional dependence:
\begin{equation}
    \frac{\sigma_{E_{\mu}}}{E_{\mu}} (\%) = \frac{a}{\sqrt{E}} \oplus \frac{b}{E} \oplus c,
    \label{eq:resolution}
\end{equation}
where \textit{a, b, c} are constants and $E_{\mu}$ is the incident particle energy in MeV. Here, $a/\sqrt{E}$ is a statistical term used to express the contribution from Poisson processes, such as photostatistics, to energy resolution. The $b/E$ term parameterizes noise contributions to energy resolution from electronics and PMTs, and the constant c parameterizes contributions from shower leakage, crystal non-uniformity, and intra-crystal miscalibrations to energy resolution. In single crystal testing, the photosensor used to read out the crystal is operated at a high voltage where noise is minimized, thereby resulting in $b\rightarrow0$ in the fit to Equation \ref{eq:resolution}. The constant term $c$ is also assumed to be dominated by crystal non-uniformity for single crystal tests. We find $a = (3.84 \pm 0.19)$ $\sqrt{\text{MeV}}$ and  $c = (0.64 \pm 0.57)$. Because the PIONEER experiment operates at a higher energy range than radioactive sources ($\mathcal{O}(10)-\mathcal{O}(100)$\,MeV), the stochastic term will be greatly suppressed in the PIONEER energy regime.


\subsection{Uniformity measurement}

\begin{figure}[H]
    \centering
    \includegraphics[width=0.8\textwidth]{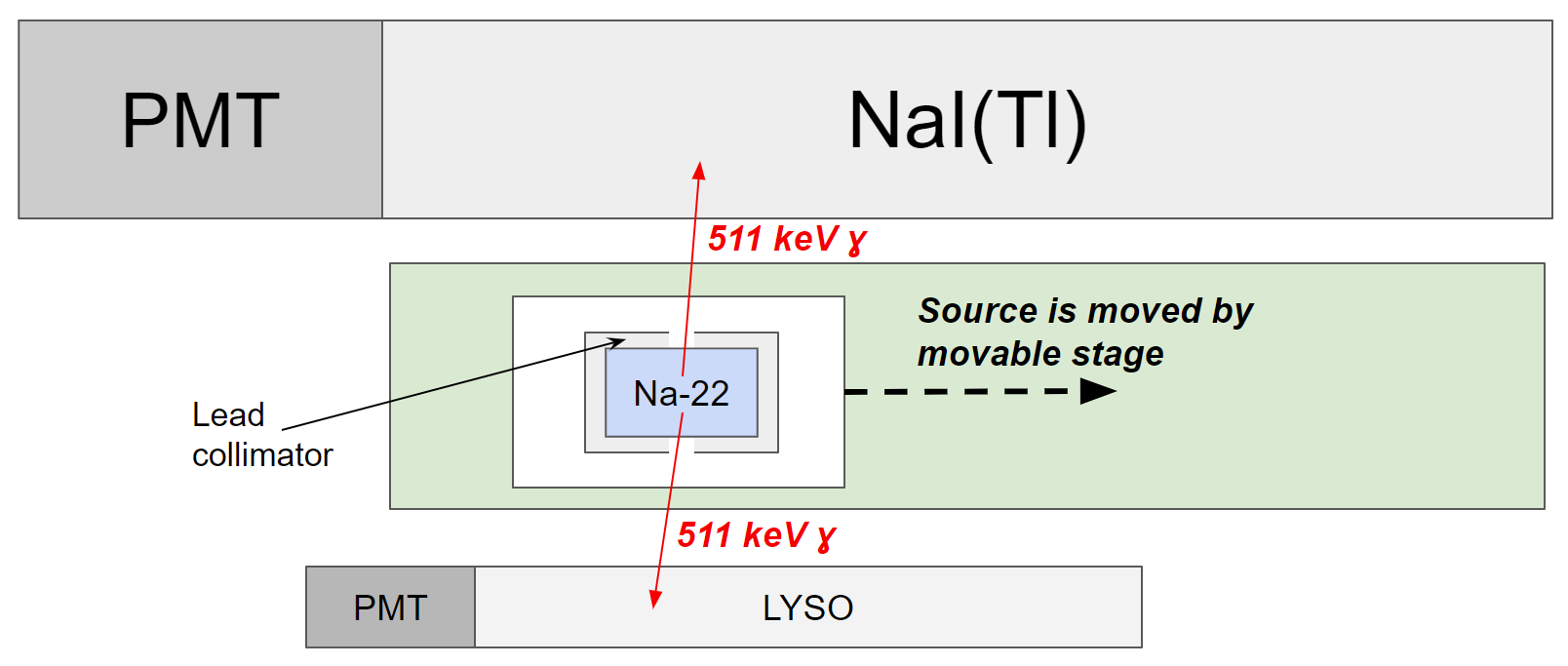}
    \caption{Schematic of the setup using 511 keV $\gamma$-rays from Na-22 to test longitudinal uniformity of a LYSO crystal. Back-to-back $\gamma$-rays from Na-22 are collimated so that a hit in a large NaI(Tl) detector corresponds to a time-coincident hit in the LYSO crystal. The Na-22 source is moved using an automated movable stage along the long axis of the LYSO crystal and the LYSO crystal response uniformity is recorded along this axis.}
    \label{fig:LongitudinalSetup}
\end{figure}

A Na-22 source producing two back-to-back \SI{511}{keV} $\gamma$-rays through $e^{+}e^{-}$ annihilation was used to measure the longitudinal response uniformity (LRU) at positions separated by \SI{1}{cm} along the length of the LYSO crystals. Measurements were made at 15 positions. 
A schematic of the tomography setup is shown in Figure \ref{fig:LongitudinalSetup}. A large NaI(Tl) detector was positioned parallel to the LYSO crystal and used as a trigger such that a 511 keV $\gamma$-ray entering the NaI(Tl) would be time coincident with a 511 keV $\gamma$-ray hitting the LYSO crystal. This trigger allowed for the source to be positioned \SI{4}{cm} from the LYSO crystal and collimated in both directions with a \SI{2.5}{cm} thick lead collimator whose hole was 1 cm in diameter; with such a setup, triggers on LYSO intrinsic radioactivity could be minimized. Waveforms from the LYSO were integrated and histogrammed in an identical manner to the procedure described in the preceding discussion of energy resolution measurements (Section 2.2). The peak position of the \SI{511}{keV} gamma was extracted from the fit of the waveform integral distribution and was tracked as a function of longitudinal position along the crystal. 

To quantify the uniformity of light intensity along the length of esra crystal, a linear fit was performed to the data. Uniformity is then reported as the 
difference in the light output obtained from the fit evaluated at the two ends of the crystal, normalized to the average light output obtained along the length of the crystal.
Relative light outputs with respect to LYSO 4 (defined to be 1.00) are defined to be the average light output of a crystal across the 15 positions in a tomography scan and are found to be 0.83-1.17 for the 10 crystals with no wrap. The LRUs of the 10 crystals with no wrap were found to be 1.8-5.2\% and the energy resolutions found as the average of the three data points taken at the center of the crystal were found to be between 5.8-7.4\% ($\sigma$/E) at 511 keV. These energy resolution measurements of the 10 crystals are presented with their manufacturer measurements in Table \ref{tab:SICCASmeasurements}.

\vskip 1 em

\begin{figure}[H]
    \centering
    \includegraphics[width=0.95\textwidth]{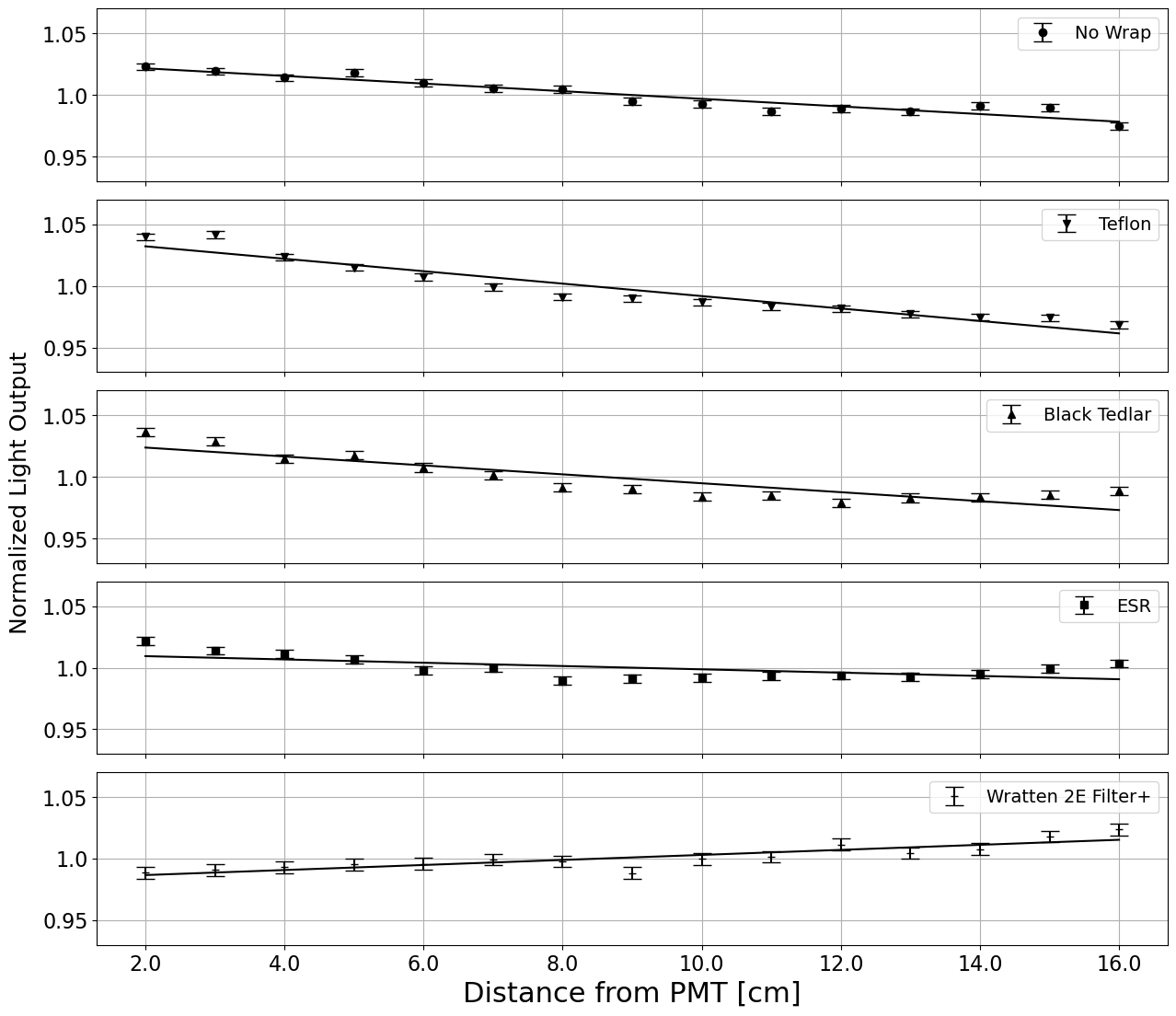}
    \caption{Variation in normalized light output for different choices of LYSO crystal wrapping and wavelength filter. Light output was determined using the peak position from a Gaussian fit of energy deposits of a 511 keV gamma from a Na-22 source positioned along the long axis of the crystal. The light outputs were normalized to the average light output along the length of the crystal.}
    \label{fig:SingleXtalTomography}
\end{figure}

Longitudinal response uniformity was measured for four additional combinations of filter and wrapping: a teflon wrapping, an ESR wrapping, a black tedlar wrapping, and a bare crystal with a Kodak Wratten 2E high-pass filter placed between the LYSO crystal and PMT. The high-pass filter was used to eliminate optical photons from scintillation with wavelengths less than 405 nm; in this wavelength regime, the absorption length of light within LYSO is shortest and can therefore introduce unwanted position dependence to the light output of the crystal. Results from LRU tests are shown in Figure~\ref{fig:SingleXtalTomography}.

Relative light outputs and LRUs for the four filter/wrapping combinations are shown in Table~\ref{tab:singleUniformities}. ESR was found to be the preferred wrapping for LYSO crystals and was ultimately used in higher energy tests due to its high light output and highly uniform response. Due to losses in light output and worsened couplings at the interfaces between the filter and PMT/crystal, the Kodak Wratten 2E high-pass filter was not used in results from later array tests presented in this study.

\begin{table}[]
    \centering
    \begin{tabular}{|c c c c|}
      \hline
      Wrapping & Filter & Relative Light Output & Non-Uniformity \\
      \hline
      None & None & \underline{1.00} & 4.3\% \\
      Teflon & None & 1.36 & 7.1\% \\
      Black Tedlar & None & 0.92 & 5.1\% \\
      ESR & None & 1.40 & 2.1\% \\
      None & Wratten 2E & 0.52 & 2.9\% \\
      \hline
    \end{tabular}
    \caption{Response uniformities along the length of LYSO crystals for different combinations of wrapping and filter. Relative light outputs are presented with respect to the case where the crystal is not wrapped and no filter is used.  }
    \label{tab:singleUniformities}
\end{table}

\section{LYSO Array Beam Test}
\label{sec:beamtests}

\subsection{LYSO calorimeter setup}

\begin{figure}[H]
    \centering
    \includegraphics[width=0.8\textwidth]{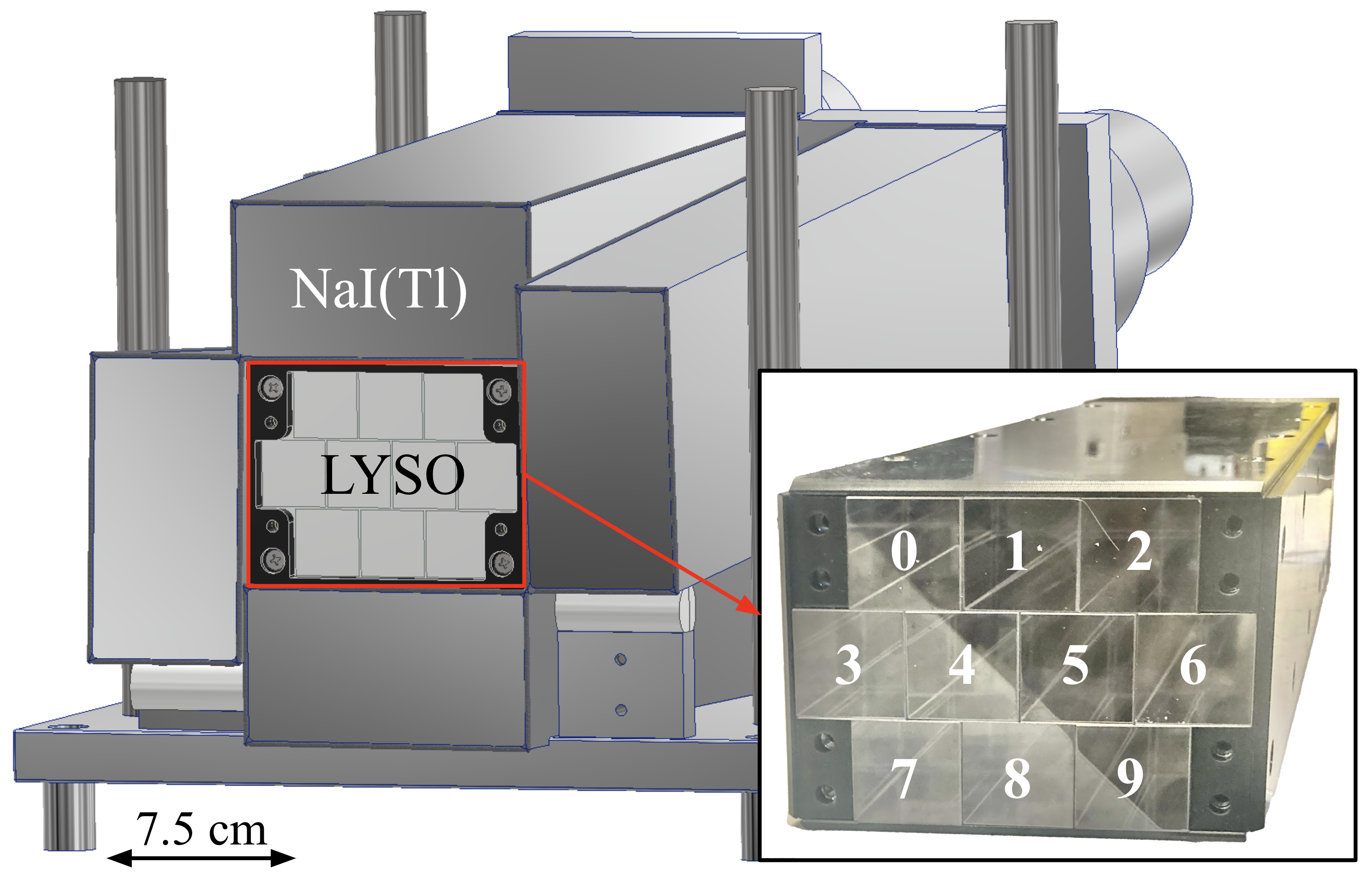}
    \caption{Front-facing image of the ten crystals forming the LYSO array. The array was surrounded by four large NaI(Tl) detectors that were used to veto events where energy leaked from the LYSO core of the detector.}
    \label{fig:LYSOarray}
\end{figure}

Once individual crystal responses were measured, an array was assembled to test responses at higher energies. In order to contain an electromagnetic shower in the lateral dimensions, 10 LYSO crystals were combined into an array as shown in Figure~\ref{fig:LYSOarray}. The crystals were arranged in 
rows of three, four, and three 
centered vertically around a common axis to ensure that the middle two crystals were surrounded on all sides. Thin pieces of ESR fitted exactly to the crystal dimensions were placed between the crystals to maximize light output and response uniformity. On the back side, each was coupled using EJ-550 optical grease to a Hammamatsu R1450 photomultiplier tube with an effective photocathode diameter of 1.5\,cm (36\% active coverage of the square face of the LYSO crystal). PMTs were operated at voltages between -750 V and -950 V to minimize PMT non-linearity effects resulting from linear PMT voltage dividers on three of the PMTs (see Section 3.10 for more discussion). Calibrations of the PMTs found a response of approximately 500 photoelectrons per MeV. A 3D-printed plastic housing was used to hold the LYSO crystals and PMTs in place and to shield them from outside light.

The LYSO array was surrounded by four large NaI(Tl) detectors housed in thin aluminum to catch the tail of the showers. The whole calorimeter setup was mounted 162.5 cm from the final beam element on an XY movable table capable of moving the detectors in a plane perpendicular to the beam with millimeter-level precision. This allowed for the beam to be focused on different crystals within the array; an automated calibration sequence was developed to move the calorimeter setup such that the beam center was aligned with the center of each of the ten LYSO crystals in the array.

Three upstream scintillator detectors at fixed locations with respect to the beam were used to trigger and monitor the beam profile. The timing of the trigger was provided by a $T0$ counter made from a $25 \times 25 \times 1$\,\si{mm^3} plastic scintillator (BC-404) that was read out by a Hamamatsu R7600U metal package photomultiplier tube operated at +780\,\si{V}. The $T0$ counter was designed to match the front area of an individual  LYSO crystal to restrict triggered events to a single crystal in the array. The second component of the trigger was a VETO detector made from $180 \times 180 \times 5$\,\si{mm^3}  scintillator with a 22\,mm diameter hole in its center. The VETO was read out by a Hamamatsu S13360-3050PE Silicon Photomultiplier (SiPM). The $T0 \cdot \overline{VETO}$ coincidence  only triggered when charged particles passed through the hole in the VETO detector. Importantly, the VETO covered most of the LYSO array and therefore prevented particles from the beam edge from triggering data acquisition, as well as pileup events of particles missing the other upstream detectors. A $12$ by $12$ channel beam hodoscope located immediately downstream of the $T0$ was used to provide precise position information for particles entering the LYSO array. The hodoscope was composed of two planes of BC-404 scintillator with an active area of $24 \times 24$\,\si{mm^2}. Each plane was composed of 12 scintillator strips of approximate dimensions $2 \times 30 \times 1$\,\si{mm^3}. The planes were oriented orthogonally to one another to provide a horizontal ($x$) and vertical ($y$) coordinate. Each strip, wrapped in teflon to improve light collection and reduce crosstalk,  was read out using a Hamamatsu S13360-3050PE SiPM operated at 54.5\,V that was attached to one end of the scintillator bar. The hodoscope was operated as a binary detector. In each plane, the hit location was associated with the center of the scintillator bar producing the largest signal, leading to a $2/\sqrt{12}$ \si{mm} position resolution.

\begin{figure}[H]
\begin{tabular}[c]{cc}
    \includegraphics[width=0.465\textwidth]{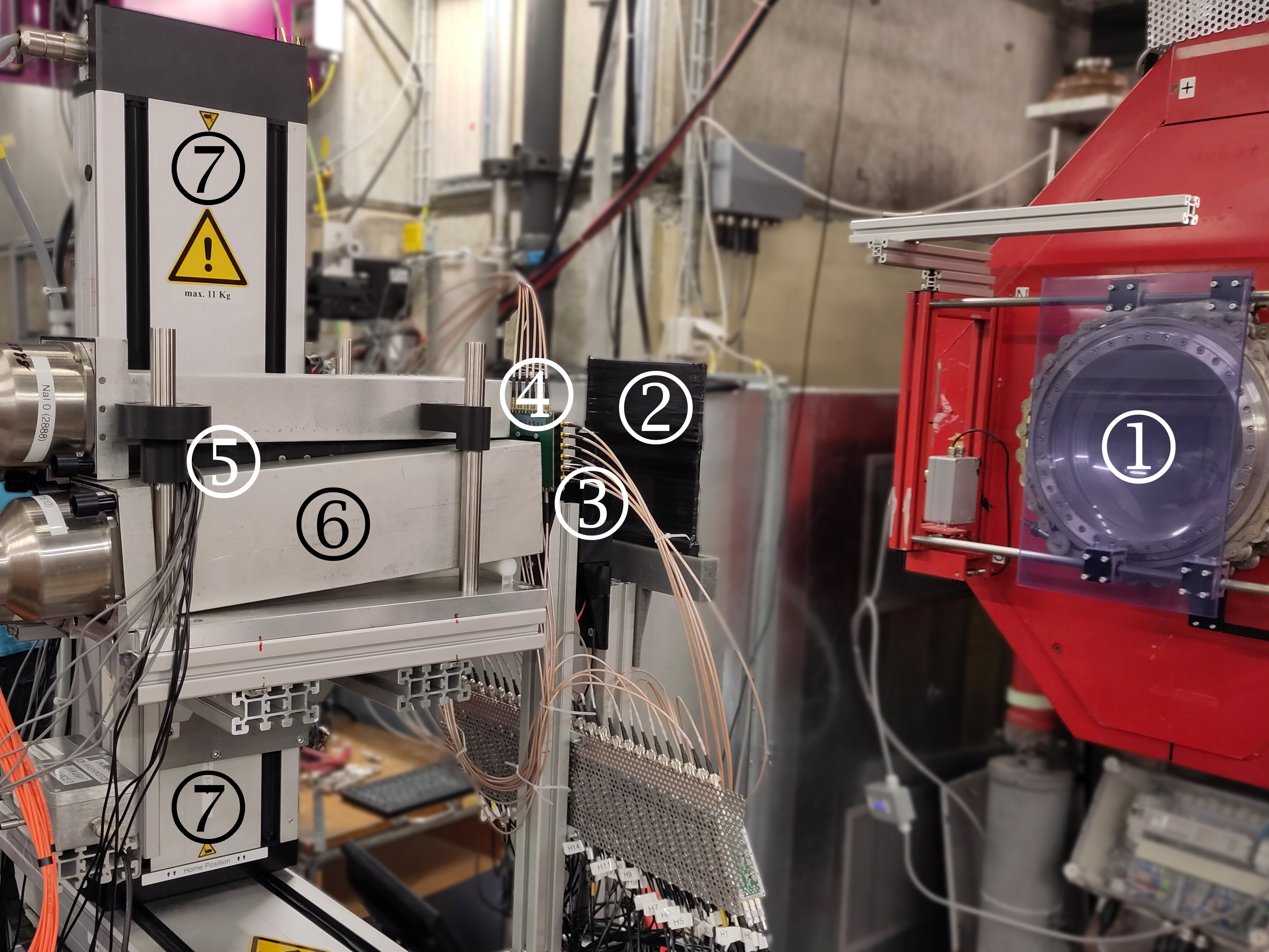}&
    \includegraphics[width=0.465\textwidth]{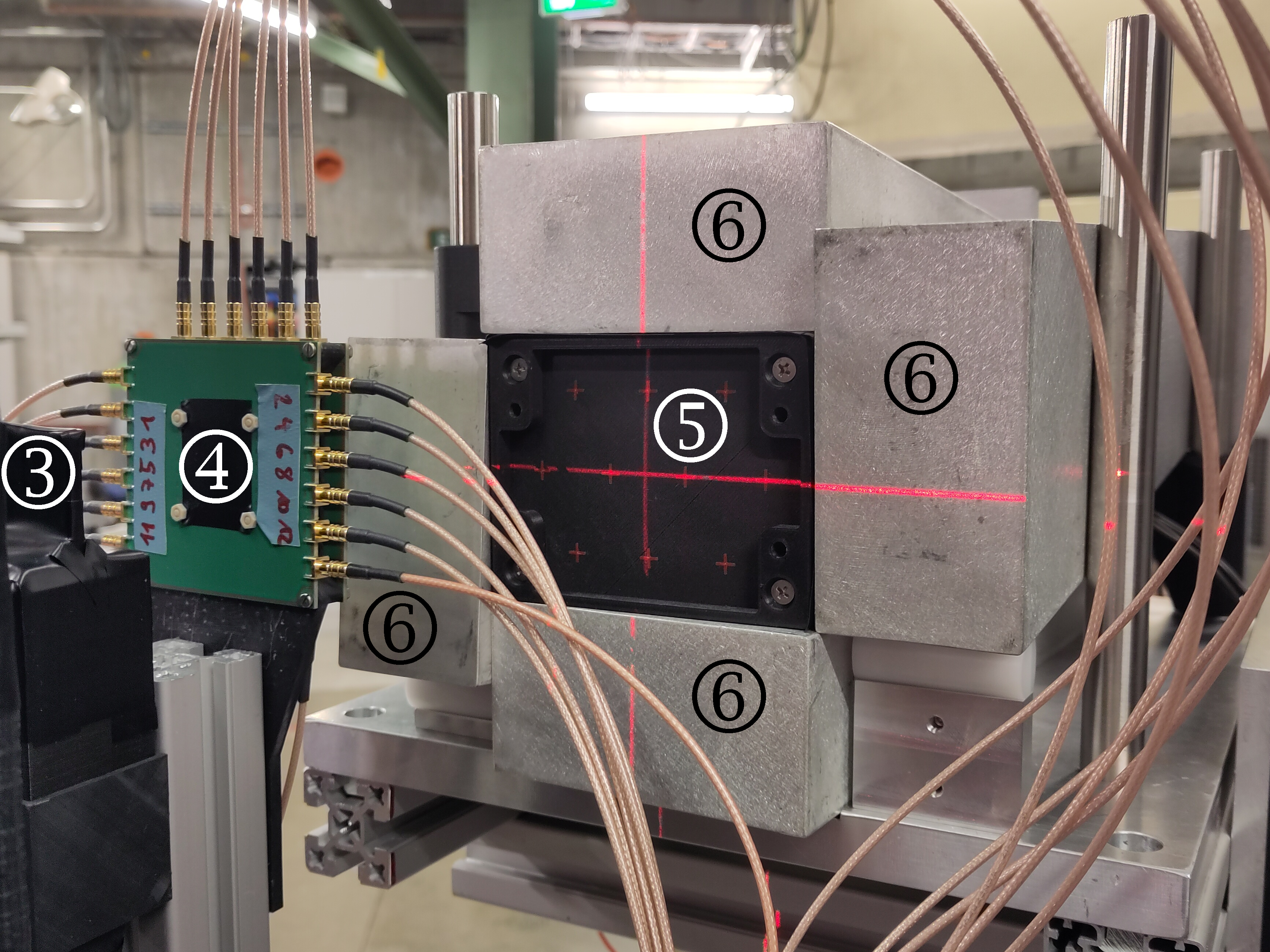}\\
  \footnotesize{(a)}&\footnotesize{(b)}\\
	\end{tabular}
	\caption{A picture of the full detector setup during the PSI test beam (a) and a close-up of the front-face of the calorimeter during laser alignment (b). Positrons coming from the last quadrupole magnet \raisebox{.5pt}{\textcircled{\raisebox{-.9pt} {1}}}
 pass the VETO counter \raisebox{.5pt}{\textcircled{\raisebox{-.9pt} {2}}}, T0 \raisebox{.5pt}{\textcircled{\raisebox{-.9pt} {3}}} and beam hodoscope \raisebox{.5pt}{\textcircled{\raisebox{-.9pt} {4}}}, before depositing their energy in the LYSO array \raisebox{.5pt}{\textcircled{\raisebox{-.9pt} {5}}}. The LYSO crystals, together with the surrounding NaI detectors \raisebox{.5pt}{\textcircled{\raisebox{-.9pt} {6}}} are mounted on a movable XY table \raisebox{.5pt}{\textcircled{\raisebox{-.9pt} {7}}}.}
	\label{fig:beamtime_setup}
\end{figure}

An LED calibration system was used to monitor voltage stability of the PMTs. At the back of each crystal, an optical fiber was positioned at the corner to shine UVA light of 365\,nm into the crystal to excite it, causing it to fluoresce. Two monitor detectors, made from a piece of plastic scintillator coupled to a PMT, received optical fibers from the same bundle as the LYSO crystals and served as external monitors of LED stability. The entire setup is shown in Figure~\ref{fig:beamtime_setup}.

\subsection{Beamline setup}

The LYSO calorimeter array was tested in November 2023 at the $\pi$M1 beam line at the Paul Scherrer Institute in Switzerland. 
Its 590\,MeV proton beam is bunched with a frequency of 50.6\,MHz. The $\pi$M1 beam line is incident on a  2\,mm thick rotating carbon target where pions are dominantly produced. Electrons (or positrons) are emitted from the target region from neutral pion production followed by pair production in the target.  They are guided along a 21\,m magnetic channel that features two 75° dipole bending magnets with an intermediate focus where momentum is dispersed. This dispersion allows for precise momentum selection by collimation, which is critical in the determination of detector energy resolution.

The beam line can be tuned for either polarity, but positively-charged particles were selected for characterization of the LYSO calorimeter, as positrons are the PIONEER signal. When properly tuned, positrons dominate the beam composition at our region of interest below 100\,MeV. Additionally, particle species can be separated in post-processing using a time-of-flight analysis to determine the differing relative particle arrival times with respect to the cyclotron radio frequency (RF), caused by their distinct velocities at the same momentum.

\subsection{Beam parameters}\label{sec:beam}
\begin{figure}[H]
    \centering
    \includegraphics[width=0.8\textwidth]{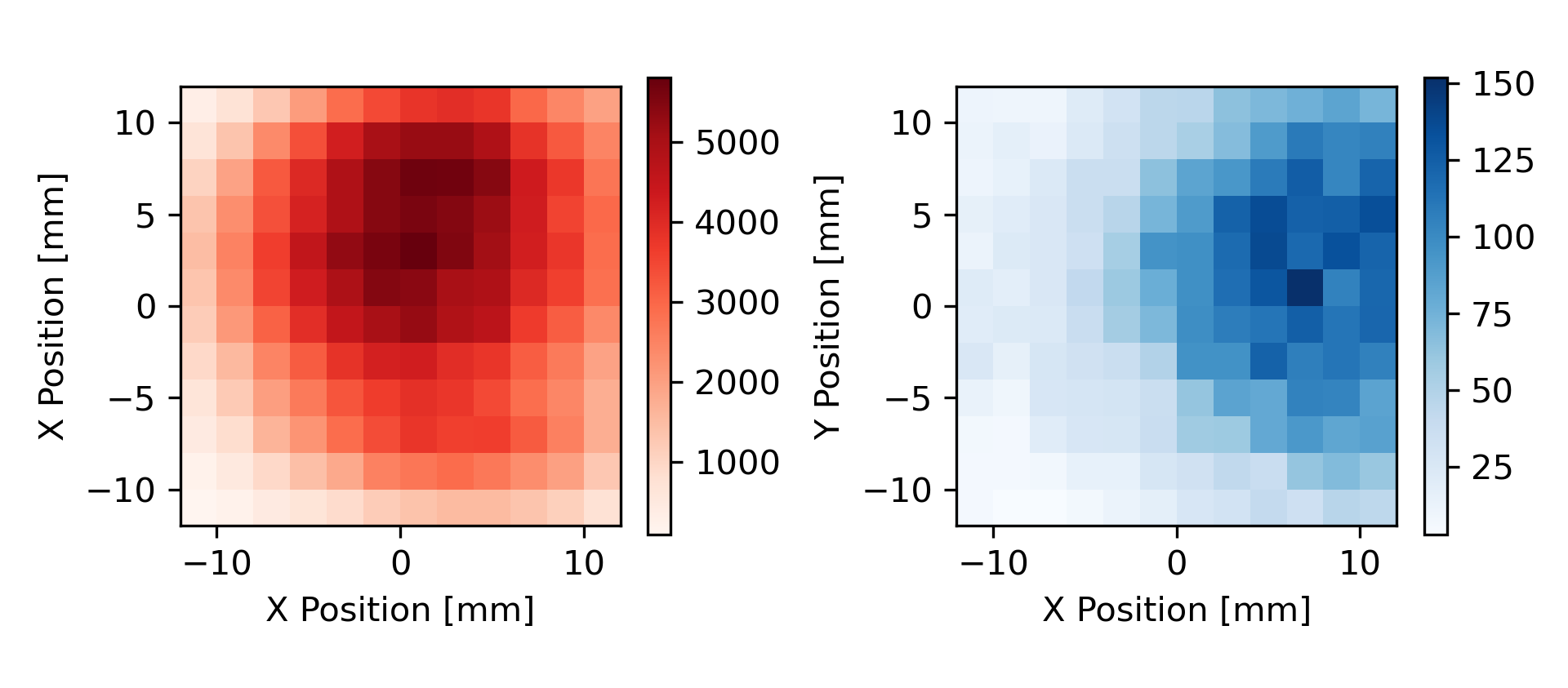}
    \caption{Beam profile measured by the hodoscope for positrons in red and muons in blue, separated using an RF cut. It can be seen that the two particle distribution centers are approximately \SI{7.5}{mm} apart.}
    \label{fig:beam_profile}
\end{figure}

The hodoscope provides a beam profile for particles hitting the LYSO array as seen for a beam momentum of 66\,MeV/$c$ in Figure~\ref{fig:beam_profile}. From the hodoscope imaging of the beam, we find a beam spot width FWHM of about 20\,mm for particles missing the VETO detector. The positron rate -- first limited by our upstream slit settings -- was typically 1.5\,kHz for momenta near 70\,MeV/$c$; an analog gate was used to throttle the data acquisition (DAQ) rate to $\sim 300$\,Hz to ensure data acquisition stability.

The beam momentum bite $\sigma_p/p$ was estimated using the timing of the events. Under the assumption that the spread in particle arrival times with respect to the RF was caused exclusively by the varying particle speeds, one can calculate $\sigma_p$ from the width of the arrival time distribution $\sigma_t$. This assumption can only be used for slower muons and pions, as the time distribution is dominated by the time resolution of our detector for highly relativistic positrons. The beam momentum bite at 70\,MeV/$c$ was determined to be less than 0.65\% $\sigma_p/p$ and to ultimately be a subdominant contribution to the reconstructed energy resolution.

\subsection{Data acquisition}
\begin{figure}[H]
    \centering
    \includegraphics[width=0.8\textwidth]{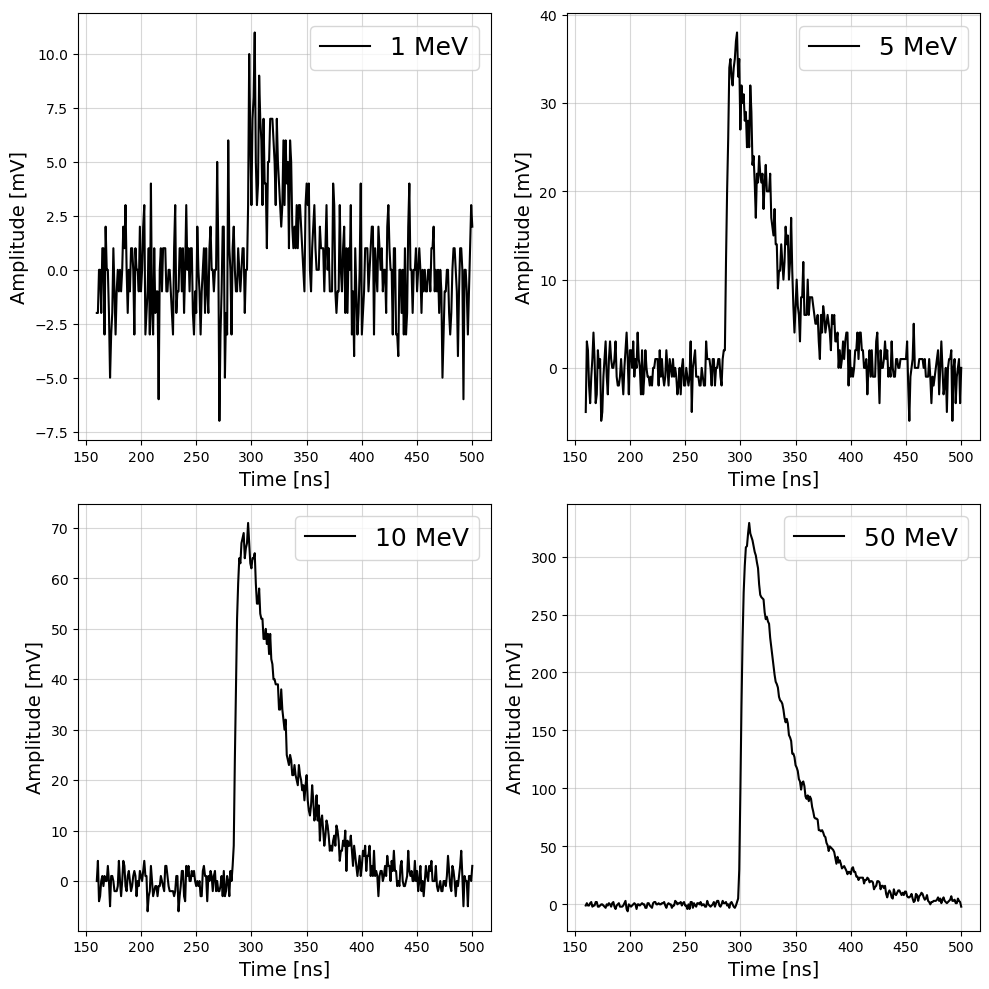}
    \caption{A representative selection of waveforms in LYSO 4 corresponding to energy deposits ranging from 1-50 MeV.}
    \label{fig:PSIwaveforms}
\end{figure}

An incoming positron that passed the $T0 \cdot \overline{VETO}$ trigger obtained an $(x,y)$ tag from the hodoscope with 69\% efficiency before it ultimately deposited its energy in the LYSO/NaI(Tl) array. A representative selection of waveforms corresponding to different energies deposited in the target LYSO crystal of the array (LYSO 4) is shown in Figure~\ref{fig:PSIwaveforms}. The collected charge was obtained for each signal waveform through integration. A pedestal average and pedestal variance was calculated from the first and last 100\,ns of the digitization window. The pedestal average with the smaller variance was then used as the final pedestal baseline. The time extracted from the crystal with the highest amplitude pulse in the array $t_{max}$ was used to set the integration window $[t_{max} - 25, t_{max} + 180]$\,(ns).


\subsection{Longitudinal response uniformity}

Longitudinal response uniformity measurements were made by directing high-energy muons through the lateral faces of crystals in the array at different positions along the LYSO crystal length. This was done by removing the NaI(Tl) detectors and rotating the array by 90 degrees. The beam momentum was increased to 210\,MeV/$c$ where the beam composition is dominated by pions and muons; muons were selected using the RF phase and energy deposition cuts. Their kinetic energy of $\sim $130\,MeV is close to minimum ionizing. Through-going muons deposit energy according to a Landau distribution peaked at approximately 23\,MeV in each of the crystals as they traverse the top row of three crystals $\sim $12.5\,mm from the crystal bottom. 

\begin{figure}[H]
    \centering
    \includegraphics[width=0.95\textwidth]{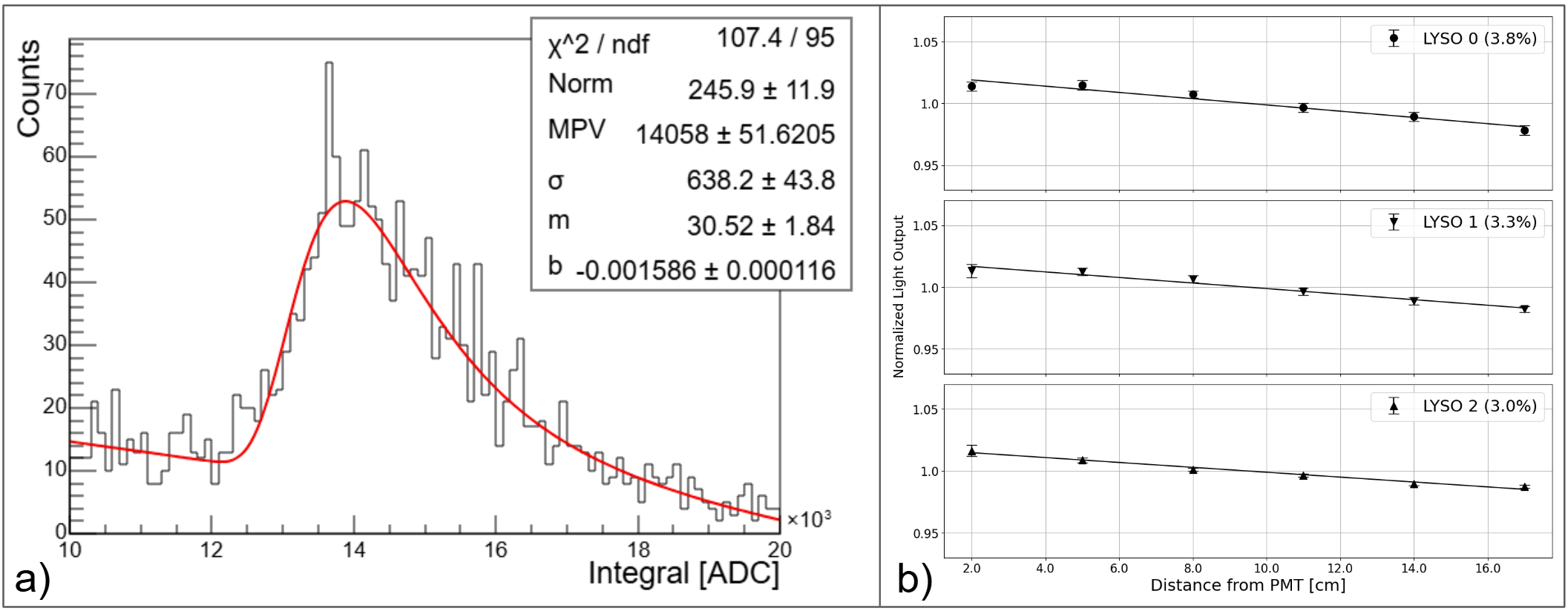}
    \caption{a) Energy deposition spectrum in a single LYSO crystal from a minimum ionizing 210 MeV/c muon penetrating its lateral face. The energy spectrum is fit using the sum of a Landau distribution used to model muon energy deposition and a linear background. A peak position is extracted from the fit; variation in this peak is tracked along the LYSO crystal length. b) Variation in light output along the longitudinal axis for three crystals in the top row of the array. This variation was found to be 3-3.8\% and is indicated for each crystal in the legend of each subplot.}
    \label{fig:muTomoRes}
\end{figure}

The trigger used in muon tomography measurements required an additional scintillator $\textit{S1}$  positioned downstream of the array. This scintillator was positioned co-linearly with the $T0$ counter such that straight particle paths through the array were selected with the coincidence $T0 \cdot \overline{VETO} \cdot S1.$ Six positions spaced 3\,cm apart along the 18\,cm length of the array were measured 
and the energy deposition was recorded for each of the three crystals. The resulting Landau distributions were fit to determine the Most Probable Value as shown in Figure \ref{fig:muTomoRes}a. This peak was tracked for each of the three crystals at each of the six positions as shown in Figure \ref{fig:muTomoRes}b. A 3-3.8\% variation in response was observed along the length of each of the three crystals, similar to Na-22 tomography measurements shown in Table \ref{tab:singleUniformities}. A 0.2\%/cm variation along the crystal length was calculated using {\tt GEANT4} simulation to contribute less than 0.25\% to energy resolution at 70 MeV.

\subsection{Calibration}
Proper reconstruction and summation of energy deposits requires a precise calibration of each PMT in the system. After building the LYSO array and coupling the crystals to their respective PMTs, low energy calibrations were used to set the PMT voltages. The voltage was set such that the peak of the intrinsic radioactivity signal at approximately \SI{0.6}{MeV} corresponded to a \SI{5}{mV} signal. The PMTs of the NaI(Tl) crystals were set to high voltages such that the \SI{1.33}{MeV} peak from Co-60 corresponded to a \SI{100}{mV} signal; this was done to maximize the ability of NaI(Tl) detectors to catch low-energy leakage from the LYSO array and veto events where leakage was present. This initial intra-crystal calibration was done to minimize the relative differences in response between crystals at higher energies by setting a rough energy scale. 

An automated calibration run was performed with positrons at beam momenta from 30 - 100\,MeV/$c$ for each crystal to normalize the relative signal output between PMTs for subsequent data taking runs. The XY table holding the crystal detector was moved such that the beam was centered on each of the ten LYSO crystals for 30,000 trigger events. For each single crystal into which the beam was aimed, energy deposits were histogrammed and fit to obtain the peak of the resulting energy deposit distribution. Simulation in {\tt GEANT4} showed that this peak corresponded to approximately 60\,MeV for a 70\,MeV/$c$ beam positron. The normalized peak energy for the single crystal was used as the calibration weight in the weighted sum of energy deposits in the LYSO array to set a high accuracy calibration between the crystals. The total deposited energy of an event in the array is then:
\begin{equation}
    E = \sum_{i=0}^{9} c_i E_i
\end{equation}
where $c_i$ are the calibration weights and $E_i$ are the energy deposits in each crystal. The NaI(Tl) detectors received additional in-situ calibrations in each data run by using the 511\,keV peak in energy deposition arising from positron-electron annihilation to calibrate between the detector responses.

\subsection{Selection Criteria}\label{sec:cuts}

Three selection criteria were used to maintain data quality at the event level for resolution measurements. The first event-level cut was an RF-phase cut using different time-of-flights of the different particle species in the beam to select positrons. Beam contamination from muons and pions was only a significant contribution at energies above \SI{70}{MeV}, and contributions from beam particles to energy resolution was essentially zero at all energies given the relatively low kinetic energy of muons/pions when compared to positrons of the same momentum. The second event-level cut was made using the T0 upstream detector to ensure a single hit in each event. A simple algorithm was used to identify T0 waveforms with multiple peaks and a cut was made whenever multiple peaks were detected to eliminate these events.

The third cut was done to maximize containment of electromagnetic showers in the 10 LYSO crystals through a removal of events with energy deposits in the NaI(Tl) detectors surrounding the LYSO array. The 1.81 and 2.42 Moliere radius semi-major axes of the LYSO array are not capable of complete shower containment in many events; often this leakage from the LYSO array consists of a small number of \SI{511}{keV} $\gamma$-rays produced by positron-electron annihilation in the final stages of a shower. Frequent low-energy leakage broadens the energy distribution peak used to obtain energy resolution and results in an unwanted geometric contribution to the energy resolution of the detector. To minimize these geometric contributions to energy resolution, an absolute veto of events with energy deposits above noise in the NaI(Tl) detectors was employed. Importantly, events with small energy deposits from Compton scattering of \SI{511}{keV} $\gamma$-rays in the NaI(Tl) were removed. Such scattering events deposited less than \SI{0.5}{MeV} in the NaI(Tl), but typically indicated energy lost from the array that would degrade resolution\footnote{Additional discussion of this veto and its effect on energy resolution is discussed in Section 3.10.}.

Determination of energy and spatial resolution was done using the aforementioned calibrated energy sum for events passing these three cuts. Additionally, timing and waveform quality conditions were employed to select the signals to be included in this sum. The crystal with the largest energy deposit was used to set the time of the shower $t_{c}$ such that other crystals with deposits above 0.5 MeV were required to have a time in the interval $[t_c - 10, t_c + 30]$\,(ns). This timing cut significantly reduced addition of pulses from the intrinsic radioactivity of LYSO into the reconstructed energy sum of the positron. The broad, low-energy spectrum of LYSO intrinsic radioactivity would worsen energy resolution if included in the reconstructed energy sum. For energy deposits less than 0.5 MeV, a Gaussian filter was applied to the corresponding signal waveform such that high frequency noise could be minimized. A Savitsky-Golay filter was then applied to smooth the output of the Gaussian filter and the filtered pulses with a peak time in the interval $[t_c - 20, t_c + 40]$\,(ns) were included in the energy sum.
Calibrated energy deposits of less than 0.1 MeV were found to be indistinguishable from PMT noise and were therefore not included in the weighted energy sum.

\subsection{Time resolution results}

\begin{figure}[H]
    \centering
    \includegraphics[width=0.75\textwidth]{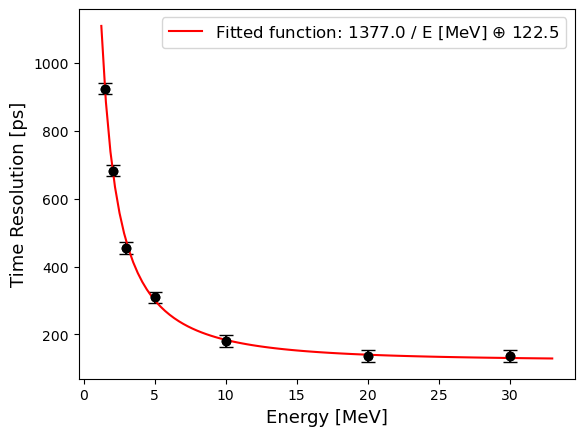}
    \caption{Dependence of time resolution of a LYSO crystal in the array on the energy deposition in the LYSO crystal for energy deposition between 1-30 MeV. A 110 ps time resolution is measured for 30 MeV energy deposits after the resolution of the reference time is removed via quadrature from $\sigma_{Fit}$.}
    \label{fig:timeResolution}
\end{figure}

A time resolution analysis was performed on datasets at a beam momentum of 70\,MeV/$c$. The time resolution analysis used the RF-phase and $T0$ cuts described in Section \ref{sec:cuts}. A containment cut was not used as multiple scattering in the LYSO array was found to have a negligible impact on the time resolution. An additional event-level cut requiring a deposit of 30\,MeV in a single LYSO crystal in the array was used to ensure a clean reference point $t_{ref}$ to which the times of other deposits in the array could be compared. 
All waveforms corresponding to energy deposits $E_{signal}$ in adjacent crystals to the reference were fit to obtain times $t_{signal}$. The differences $\Delta t =  t_{ref} - t_{signal}$ were computed and histogrammed with their associated $E_{signal}$. The standard deviation of the distributions of $\Delta t$ were determined and a time resolution was extracted for energies ranging from 1 to 30\,\si{MeV} after removing the resolution of the reference time (110 ps for 30 MeV references) in quadrature. The time resolutions were fit as a function of energy to the expression:
\begin{equation}
    \Delta t = \frac{a_{t}}{\sqrt{E}} \oplus \frac{b_{t}}{E} \oplus c_{t}.
\end{equation} From this fit, we found the fit parameters $a_{t} = (443 \pm 51)$ $\sqrt{\text{MeV}}$ $\cdot$ ps, $b_{t} = (952 \pm 63)$ MeV $\cdot$ ps, and $c_{t} = (74 \pm 16)$ ps. The dependence of time resolution on energy is shown in Figure\,\ref{fig:timeResolution}. A time resolution of \SI{880}{ps} was found at \SI{1}{MeV} and a resolution of \SI{110}{ps} was found at \SI{30}{MeV}.

A similar crystal-to-crystal time resolution analysis was performed using the LED monitoring system to simultaneously induce the scintillation mechanism of each crystal. This produced waveforms larger than 1\,V in each crystal; at this signal size, noise and stochastic contributions were greatly suppressed. In this case, the time extracted from each waveform was compared to the time in LYSO 0 which acted as $t_{ref}$ in this case. The timing resolutions of the crystals ranged from 60-80 ps.
\subsection{Position resolution and lateral uniformity}

\begin{figure}[H]
    \centering
    \includegraphics[width=0.8\textwidth]{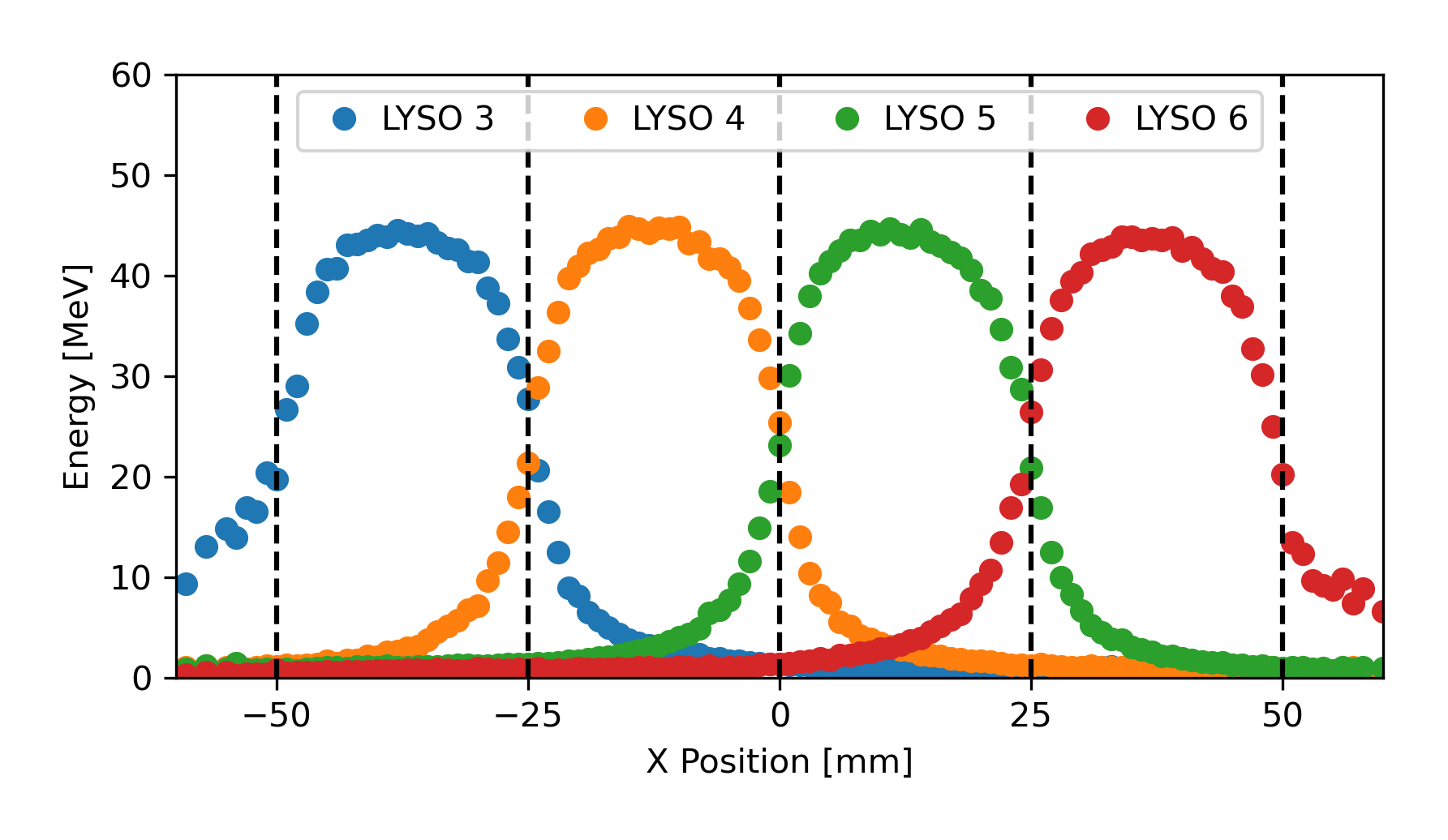}
    \caption{
    Plot of the average energy in each crystal of the center row for particles hitting each x-position along the front face of the array. The dashed lines indicate crystal boundaries, corresponding to the points where two neighboring crystals produce the same average integral.
    }
    \label{fig:position_calibration}
\end{figure}

\begin{figure}[H]
	\begin{tabular}[c]{cc}
		\includegraphics[height=0.315\textwidth]{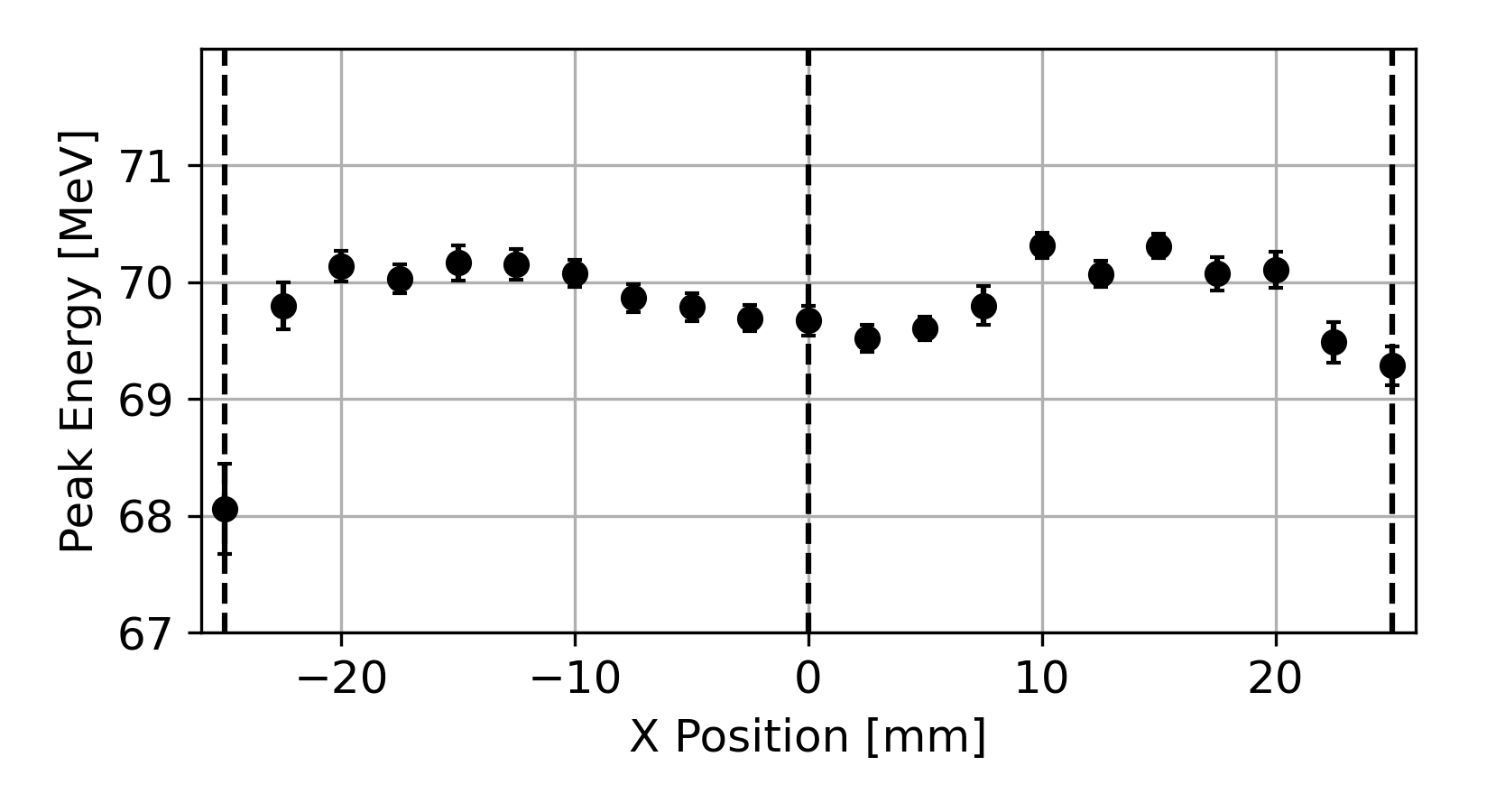}&
		\includegraphics[height=0.315\textwidth]{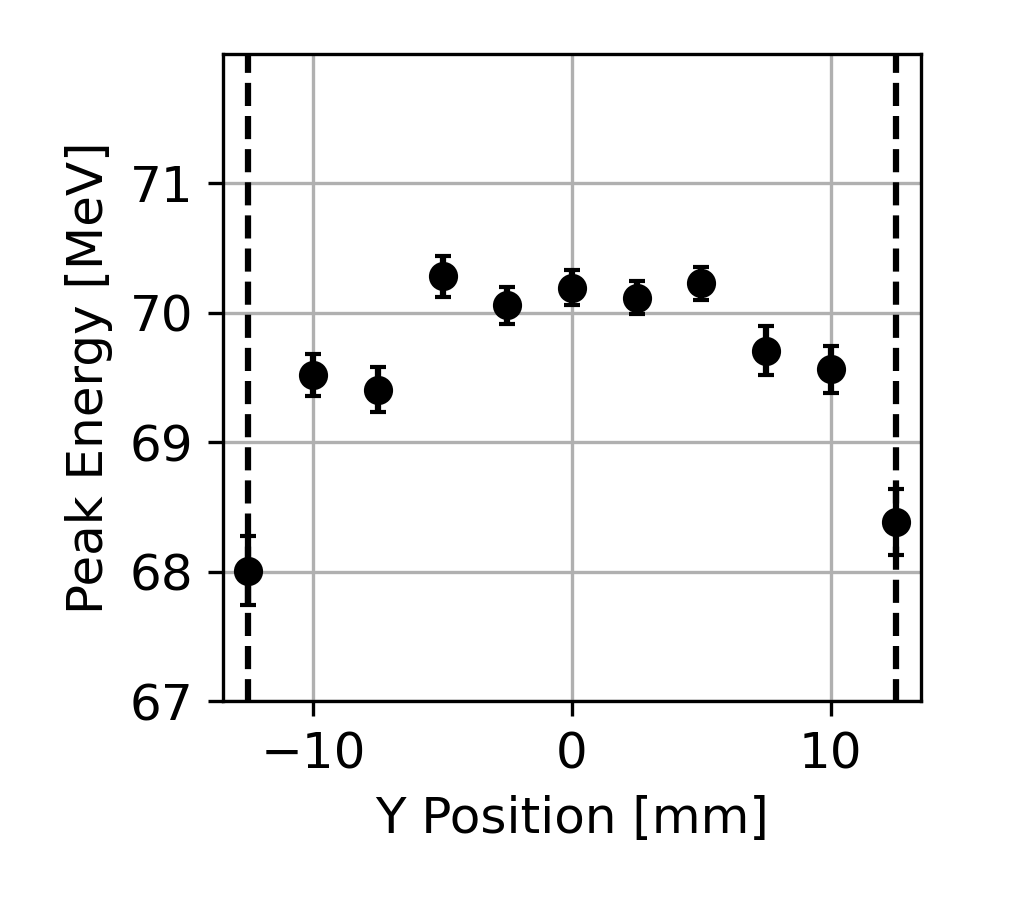}\\
  \footnotesize{(a)}&\footnotesize{(b)}\\
	\end{tabular}
	\caption{Fine scan along the horizontal (a) and vertical (b) axis of the crystal. Data was reorganized into \SI{2.5}{mm} bins to increase statistics such that the reconstructed peak energy could be obtained from a fit to the dataset. The relative change in peak energy within the center crystals is less than \SI{0.4}{\%} in the horizontal scan and \SI{0.5}{\%} in the vertical scan. The dashed lines indicate the boundaries between crystals. Peak energy decreases for points in the fine scan near the outer parts of the array due to lateral leakage.
    }
	\label{fig:fine_scan}
\end{figure}

In order to check the uniformity of the LYSO array response along the x-coordinate, datasets were taken along the x-axis in the vertical center of the LYSO array with the beam positioned in intervals of \SI{5}{mm}. A similar scan was performed along the y-axis at x = \SI{12.5}{mm}, aligned through the center of LYSO 5. Both scans were performed at a beam momentum of \SI{70}{MeV}/c.
The width of the positron beam (20 mm FWHM as described in Section 3.3) is comparable to the size of the front face of a crystal; using hodoscope information, the data can be binned according to particle hit position along the array rather than using the position of the beam center. The relative position of the hodoscope with respect to the array can be precisely calibrated using the average energy deposited in the LYSO crystals for every hodoscope channel, and relating the crystal boundaries to the points where the energy is shared to equal amounts by two neighboring crystals.  
A combined plot of all x-scan data sets showing the result of the position calibration is displayed in Figure \ref{fig:position_calibration}. 
The energy deposited in the array is expected to decrease near crystal boundaries due to particles traveling along spaces between crystals, limiting lateral uniformity of the array. In order to quantify this effect, the scan data was binned along the dimension of interest, while no cut was applied in the perpendicular direction. Then, at each location, the calibrated sum of the energies deposited in all LYSO crystals was fitted with a Crystal Ball function as described in Section \ref{sec:energy_resolution}. The resulting reconstructed peak energy as a function of the position is shown in Figure \ref{fig:fine_scan} a) for the x-axis and b) for the y-axis. The relative change in peak energy $\sigma_E / \bar{E}$ within the two center crystals was found to be less than \SI{0.5}{\%}.

Using the hodoscope position information, one can also calculate the position resolution of the LYSO array itself. Because of the relatively fine segmentation of the LYSO array, a shower typically left a signal in five to six crystals. The position of an event can be estimated through calculation of its energy-weighted center-of-mass, according to the formula found in \cite{AWES1992130}:
\begin{equation}
    (x,y) = \left( \frac{\sum_i w_i \cdot x_i}{\sum_i w_i}, \frac{\sum_j w_j \cdot y_j}{\sum_j w_j} \right),
\end{equation}where
\begin{equation}
    w_i = \mathrm{max} \left\{ 0, \left( w_0 + \mathrm{log} \frac{E_i}{\sum_i E_i}  \right)  \right\}
\end{equation}
is a logarithmic weight that  takes into account the exponential falloff of the energy deposition in the lateral direction. The free parameter $w_0$ was optimized for the specific calorimeter geometry using simulation in {\tt GEANT4}. As an alternative, a fully connected feed forward convolutional neural network with two hidden layers was trained to predict the event location using the same simulation dataset. Applying the two algorithms to real data and comparing the result to the measured position with the hodoscope revealed a position resolution of about \SI{6.0}{mm} in $x$ and $y$ for the neural network, and about \SI{6.4}{mm} for the center-of-mass algorithm. 

\subsection{Energy resolution of the LYSO array}\label{sec:energy_resolution}
\begin{figure}[h]
    \centering
    \includegraphics[width=0.8\textwidth]{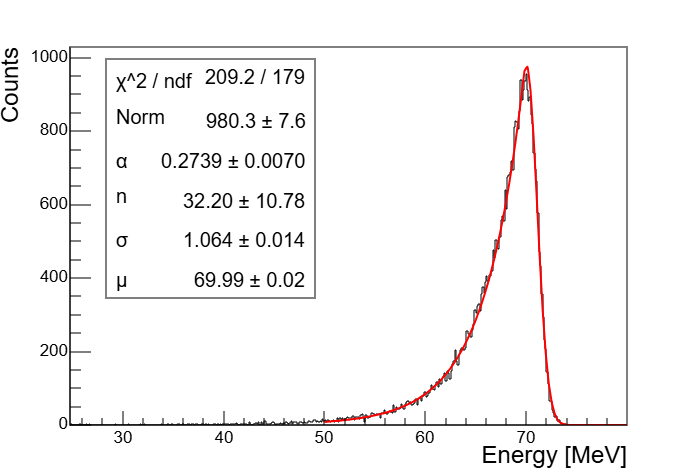}
    \caption{Summed energy deposit spectrum for a 70 MeV positron in the LYSO array. The energy resolution at 70 MeV obtained from a Crystal Ball fit is 1.52 $\pm$ 0.03\%.}
    \label{fig:Eres70MeV}
\end{figure}

The energy deposit distributions obtained from the weighted sums of crystal energies passing the event selection cuts were fit to obtain energy resolutions at each beam momentum. The fitted energy deposit distribution for a beam momentum of 70 MeV/c is shown in Figure \ref{fig:Eres70MeV}. The fit of the energy deposition distribution was done using the Crystal Ball function\footnote{The Crystal Ball function is not an exact fit of the recorded energy deposit distribution because the peak width of the distribution is not entirely due to random processes. In reality, at least two 511 keV $\gamma$ rays were produced in every electromagnetic shower where the beam positron stopped in our detector. These 511 keV $\gamma$-rays have large absorption lengths in LYSO because photoabsorption and Rayleigh scattering cross sections are marginal at this energy, and pair production is not kinematically allowed. Thus, the 1.81 and 2.42 Moliere radius semi-major axes of the LYSO detector and surrounding NaI(Tl) did not contain 511 keV $\gamma$ rays in many events, resulting in three photopeaks corresponding to whether 0, 1, or 2 of the 511 keV $\gamma$ rays were absorbed in an event. Given the 511 keV separation of the peaks and their varied heights, these three photopeaks cannot be resolved and appear as a single broadened peak that is slightly non-Gaussian.} which consists of a Gaussian core (mean $\mu$ and standard deviation $\sigma$) and power law tail (exponent $n$ and transition energy $\alpha$) to account for processes resulting in high energy loss. The standard deviation and peak energy are extracted from the fit to obtain the energy resolution: $\sigma / \mu$. At 70 MeV, an energy resolution of 1.52\% is measured using a fit to the Crystal Ball function; this resolution becomes 1.71\% when using FWHM / 2.355 due to leakage of the shower from the array. This energy resolution was unchanged when a data set was selected around the crystal boundary between the centers of crystals 4 and 5 in the center of the array with hodoscope cuts. 
In the case where the NaI detectors are not used to veto events and instead the energy deposited in NaI is included in the total energy sum, the measured energy resolution is 1.75\%. This degradation without the use of the NaI veto is reproduced in simulation. 
It arises from shower development fluctuations near the edge of the array where small amounts of energy escapes, and from Compton scattering of 511 keV $\gamma$ rays in the NaI detectors where the $\gamma$ ray deposits very little energy in the NaI before escaping the array.
Simulation in {\tt GEANT4} of a larger 7\,x\,7 array of LYSO crystals (crystal size of 2.5\,x\,2.5\,x\,18\,cm$^3$) without NaI detectors finds that the NaI veto properly reproduces the energy resolution of this larger array where lateral leakage is minimized. The results of these cases are summarized in Table \ref{tab:veto}.

\begin{table}[H]
    \centering
    \begin{tabular}{|c | c|}
      \hline
      Case & Energy Resolution at 70 MeV [\%] \\
      \hline
      Veto Used [Data] & 1.52 \\
      Veto Used [Simulation] & 1.56 \\
      No Veto [Data] & 1.75 \\
      No Veto [Simulation] & 1.71 \\
      7x7 Array [Simulation] & 1.55 \\
      \hline
    \end{tabular}
    \caption{Comparison of the energy resolutions obtained through fits to the Crystal Ball function for various configurations in data and simulation. We find that the use of the NaI detectors as vetos reproduces the energy resolution obtained for a larger array of LYSO crystals in {\tt GEANT4} simulation.}
    \label{tab:veto}
\end{table}



The energy resolution was measured for each beam momentum from 30-100 MeV/c. A fit of the energy resolution was performed using Equation \ref{eq:resolution}. In the fit, the parameter $a$ was found to converge to 0 and so a second fit without the $a$ term was performed. The fit is expressed in the form\footnote{The fit parameter $a$ used in single crystal testing was not used in this fit as different PMTs used.}:
\begin{equation}
    \frac{\Delta E}{E}(\%) = \frac{b_{a}}{E} \oplus c_{a}.
\end{equation}
We find $b_{a} = (72.58 \pm 0.54)\, \text{MeV}$ and $c_{a} = 1.16 \pm 0.01$.
The parameter $b_a$ was dominated by the photosensor noise present during testing and we expect that this noise contribution could be reduced by using PMTs operated at higher voltages. The constant parameter $c_a$ is attributed to miscalibration between the crystals, voltage instabilities, imperfect reconstruction of intrinsic LYSO radioactivity and lateral leakage from the LYSO array that was not captured in the NaI(Tl) detectors. 

\begin{figure}[H]
    \centering
    \includegraphics[width=0.7\textwidth]{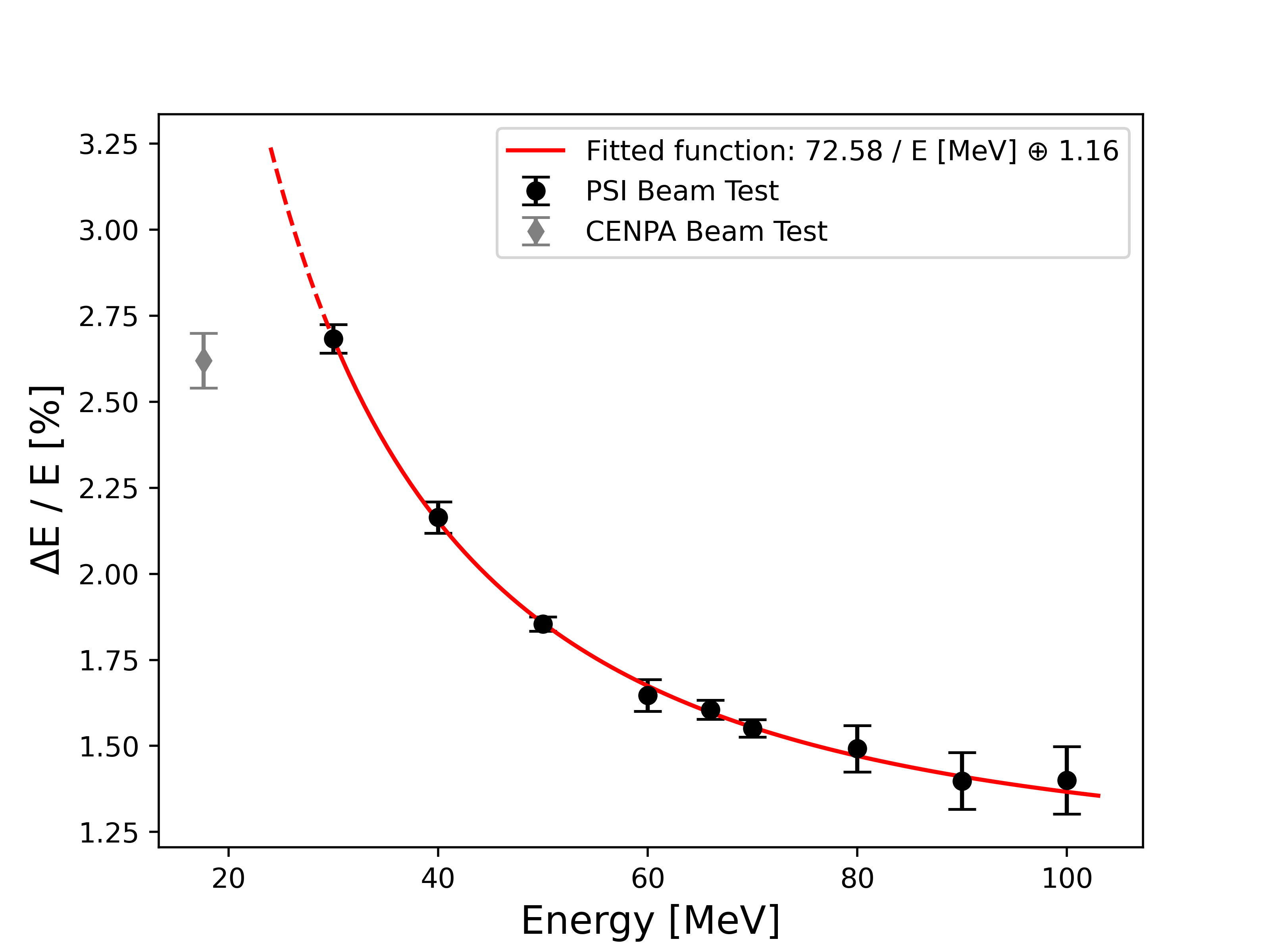}
    \caption{Dependence of energy resolution of the LYSO array on postitron energy for PSI runs and the 17.6 MeV $\gamma$ run at CENPA. Energy resolutions from PSI testing were computed from data runs where the beam was centered on LYSO 4 of the array. PMTs were operated at 300 V higher voltages in the CENPA beam test, which is expected to have significantly improved data quality and energy resolution.}
    \label{fig:resCENPA}
\end{figure}

\begin{figure}[H]
    \centering
    \includegraphics[width=0.6\textwidth]{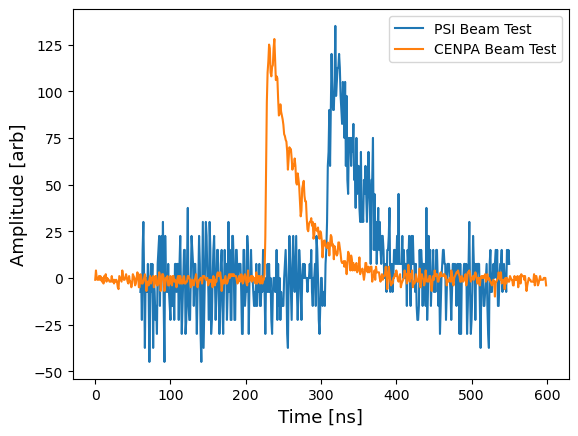}
    \caption{Two 2-MeV waveforms amplitude-normalized for differences in gain from the same LYSO crystal and PMT with different operating voltages. The PMT was operated at -1275\,V during the CENPA test beam and -975\,V at the PSI test beam, which resulted in significantly more noise in the data from the PSI test beam.}
    \label{fig:WFsPSIvsCENPA}
\end{figure}

The large noise contribution is expected to be the result of operating PMTs at the low end of their operational voltages. This was done to minimize the effects of the non-linear response from PMTs with linear voltage dividers,\footnote{The resistance at the last six dynodes is equal; this voltage divider is typically used to optimize phototube response for low light signals.} but it also increased PMT noise. An energy resolution measurement of the LYSO array using a proton-Lithium resonance was performed at CENPA after PSI testing; importantly, all PMTs had tapered voltage dividers\footnote{The relative resistances at the last six dynodes were 0.425, 0.567, 0.683, 1.00, 1.00, 1.00; tapered voltage dividers are typically used to optimize PMT linearity over a large dynamic range.} for this test and the PMTs were operated at an average of 300 V higher than at PSI. Operating the PMTs at higher voltages greatly improved the resolution obtained by the LYSO array at low energies as can be seen in Figure \ref{fig:resCENPA}. Representative 2 MeV pulses from the two different operating voltages normalized for the difference in gain are shown in Figure \ref{fig:WFsPSIvsCENPA}. Significantly more noise was found in the waveforms recorded at PSI than those at CENPA from p-Li data.  The noisier waveforms led to a degraded effective energy resolution. The CENPA test is described in more detail in Section 4.

\section{CENPA $\gamma$-ray Test}

\subsection{Test beam setup}

\begin{figure}[H]
    \centering
    \includegraphics[width=0.7\textwidth]{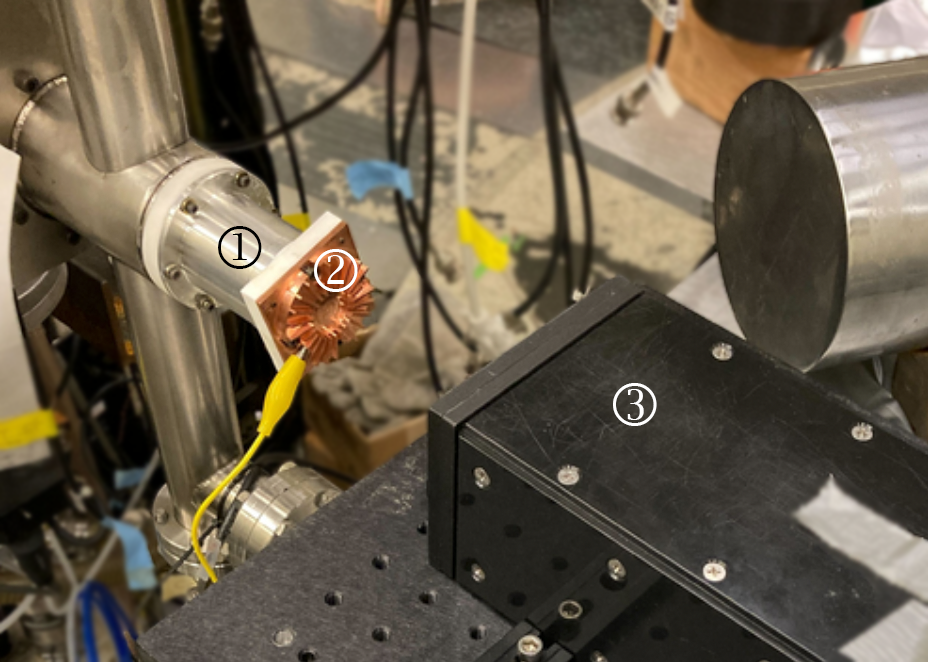}
    \caption{Picture of the CENPA test setup. The proton beam  \raisebox{.5pt}{\textcircled{\raisebox{-.9pt} {1}}} is stopped in a LiF target \raisebox{.5pt}{\textcircled{\raisebox{-.9pt} {2}}} used to produce 17.6\,MeV $\gamma$-rays. The ten crystal LYSO array \raisebox{.5pt}{\textcircled{\raisebox{-.9pt} {3}}} is positioned 10.8\,cm from the LiF target. Signals above 8 MeV in the center two crystals were used to trigger the data acquisition.}
    \label{fig:CENPAsetup}
\end{figure}

Following the tests of the LYSO detector at PSI, additional tests of the LYSO crystals were conducted at CENPA in April 2024 using a proton-Lithium reaction. In this beam test, the same LYSO array assembly was used as described in Section 3.1, but with its upstream and NaI detectors removed. The experimental setup in Cave 1 at CENPA can be seen in Figure \ref{fig:CENPAsetup}. As described in Section 3.10, the PMT bases used in the PSI tests were replaced with bases containing tapered voltage dividers that could be operated at higher voltages. The voltages were set between 1200-1300\,V so that the \SI{0.6}{MeV} peak in the LYSO intrinsic radioactivity spectrum produced a \SI{30}{mV} signal. The signals from the center two crystals were discriminated at \SI{8}{MeV} to trigger the data acquisition, and avoids gamma rays produced with energies below \SI{7}{MeV} by fluorine proton capture reactions in the target.

\subsection{The p-Li reaction}
A proton-Lithium resonance reaction has been used in calorimeter calibration for the MEG II experiment at PSI and the L3 experiment at CERN \cite{MEGII:2023fog,L3}. The reaction would also be an ideal physics channel for calibration of a LYSO calorimeter for the PIONEER experiment. The $^{7}$Li$(p,\gamma)$ reaction begins with a 440\,keV proton capture on Li-7 to produce an excited Be-8 state. This Be-8 state exists at a 17.6\,MeV level that will subsequently de-excite to the ground state or a broad level at 3\,MeV, producing 17.6\,MeV $\gamma$-rays and a broad spectrum of $\gamma$-rays peaked at 14.6\,MeV, respectively. Proton energies must be kept below 1.03\,MeV to avoid excitation of Be-8 to the 18.2\,MeV level, which produces $\gamma$-rays peaked at 16.6 and 16.9\,MeV with a broad distribution. \\
For this measurement, a tandem Van de Graaff accelerator\footnote{High Voltage Engineering Corporation Model FN} 
located at CENPA was used to accelerate protons to 1.4\,MeV with a beam current of 1\,$\mu$A. The protons were then degraded by a 7.5 $\mu$m tantalum foil before reaching a 1\,mm LiF target. Simulations done in TRIM find that the proton energy reaching the target is peaked at 550\,keV after degradation with all proton energies below the 1.03\,MeV threshold. Protons travel approximately 5\,$\mu$m in the LiF crystal before capturing and ultimately producing a roughly isotropic distribution of $\gamma$-rays.

\subsection{Data analysis and results}
\begin{figure}[H]
    \centering
    \includegraphics[width=0.7\textwidth]{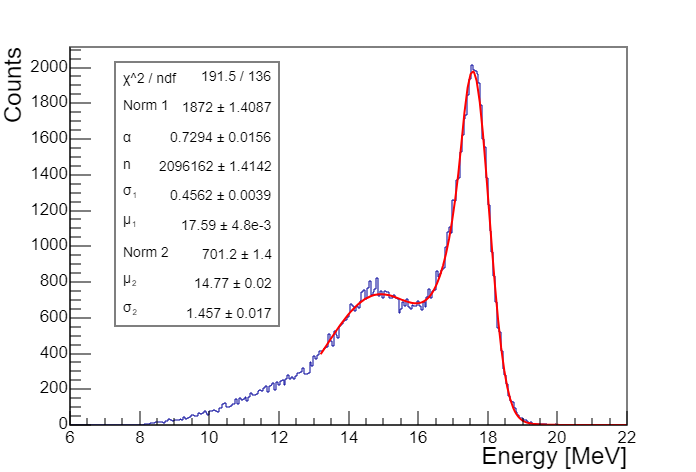}
    \caption{Energy deposit distribution for 14.6\,MeV and 17.6\,MeV $\gamma$-rays in the 10 LYSO crystal array. The distribution was fit with a sum of a Gaussian distribution (for the broad resonance at 14.6\,MeV) and a Crystal Ball function for the 17.6\,MeV $\gamma$-ray. An energy resolution of 2.6\% was measured at 17.6\,MeV. }
    \label{fig:CENPAresolution}
\end{figure}

Intra-crystal calibrations were determined in separate systematic runs where each of the crystals was discriminated and allowed to trigger the system. The peak corresponding to a 17.6\,MeV gamma in the single crystal energy deposit spectra was fit to obtain peak position and calibrate between the crystals. After calibration, 511\,keV peaks in all crystals were well aligned. The full energy deposition in the array was calculated using a weighted sum of crystal energies where the weights corresponded to the relative peak position of the 17.6\,MeV gamma peak in the single crystal. Single crystal energy depositions below 0.2 MeV were not included in the weighted energy sum. Additional timing cuts were applied that required pulses to be within a 50-ns gate defined as $[t-10,t+40]$\,ns where $t$ is the hit time of the crystal triggering the DAQ. The summed energy deposit distribution of the array is shown in Figure \ref{fig:CENPAresolution}. The distribution was fit with using a Crystal Ball function to model the sharp 17.6\,MeV resonance and a Gaussian to model the broad resonance when Be-8 de-excited to the 3\,MeV level rather than the ground state. An energy resolution of 2.6\% was measured at 17.6\,MeV. 

\section{Conclusions}
The main results of our studies on ten rectangular-shaped, next-generation LYSO crystals are summarized:
\begin{itemize}
\item The longitudinal response uniformity of the tested crystals, measured with both bench sources and muon tomography, is better than 4\%. The measurement does not depend on which end of the crystal is viewed by the photosensor.
\item With optimized photomultiplier voltage dividers, an energy resolution of 2.6\% for 17.6\,MeV gammas from a p-Li reaction is measured.
\item The timing resolution vs. energy (Fig.\ref{fig:timeResolution}) is better than 200\,ps for energies above $\sim 10$\,MeV and is 110\,ps at 30 MeV.
\item Spatial resolution from energy sharing at 70\,MeV is $\sim 6$\,mm.
\item Energy resolution dependence on energy is presented in Fig.\ref{fig:resCENPA}; for 70\,MeV positrons, a 1.5\% response is achieved. An improved resolution is expected for tests done with PMTs using tapered voltage dividers compared to the fit shown in Fig.\ref{fig:resCENPA}, especially at lower energies.
\end{itemize}

While these performance characteristics meet the goals of PIONEER, the final PIONEER calorimeter geometry requires a quasi-spherical arrangement of tapered pentagonal and hexagonal crystals to form a compact and hermetic array.  We have ordered samples of full-size, 19-$X_0$ deep crystals in several shapes and will begin studies in the near term.  

We have shown that large-format, high-density, non-hygroscopic, radiation-hard LYSO crystals, with their intrinsic high light yield and 40\,ns scintillating decay time, can serve as excellent choices for electromagnetic calorimeters in nuclear and particle physics experiments.

\section{Acknowledgments}
This work was supported by the United States Department of Energy grant DE-FG02-04ER41286 and DE‐FG02‐97ER41020. J. Liu is supported by the Hongwen Foundation in Hong Kong and the New Cornerstone Science Foundation in China. B. Davis-Purcell is supported by the Natural Sciences and Engineering Research Council of Canada. We thank the CENPA Technical and Accelerator staff for their assistance in this measurement and the PSI accelerator staff for support and delivery of reliable beam at multiple energies.  
The research institute SICCAS has been exceptionally instrumental in the development of LYSO as a scintillating option and in working with our group for customized crystals. We thank R. Gilman and E. Cline from the MUSE collaboration for sharing information about the $\pi$M1 beamline.
Expert advise from Ren-Yuan Zhu (Caltech) guided the early stages of this study. We also thank our PIONEER colleagues for their support and feedback.

\appendix
\section{Manufacturer LYSO Measurements}

Manufacturer measurements of energy resolution using the 0.662 MeV gamma produced by a Cs-137 source are compared to bare crystal measurements done at CENPA using the 0.511 MeV gamma from a Na-22 source in Table \ref{tab:SICCASmeasurements}. Energy resolutions quoted are $\sigma/E$.
\begin{table}[H]
    \centering
    \begin{tabular}{|c c c c|}
      \hline
      LYSO Number & Serial Number & $E_{Res}$ [0.662 MeV] &  $E_{Res}$ [0.511 MeV] \\
      \hline
      0 & LYSO20230519-1 & 4.5\% & 6.1\% \\
      1 & LYSO20230608-1 & 5.2\% & 5.8\% \\
      2 & LYSO20230608-2 & 4.9\% & 5.9\% \\
      3 & LYSO20230608-3 & 4.8\% & 5.9\% \\
      4 & LYSO20230608-4 & 5.2\% & 6.0\% \\
      5 & LYSO20230608-5 & 5.1\% & 7.1\% \\
      6 & LYSO20230608-6 & 5.4\% & 6.3\% \\
      7 & LYSO20230608-7 & 5.0\% & 6.2\% \\
      8 & LYSO20230608-8 & 4.7\% & 5.9\% \\
      9 & SIC220818-3 & 6.1\% & 7.4\% \\

      \hline
    \end{tabular}
    \caption{Comparison of manufacturer measurements of energy resolution for the 10 LYSO crystals at 0.662 MeV and measurements done at CENPA at 0.511 MeV.}
    \label{tab:SICCASmeasurements}
\end{table}

\bibliography{LYSO-CALO.bib}

\end{document}